\documentclass{emulateapj}

\usepackage{graphicx,hyperref,amsmath}
\hypersetup{colorlinks,citecolor=black,filecolor=black,linkcolor=green,urlcolor=magenta,pdftex}
\usepackage{epstopdf}
\usepackage{natbib}
\usepackage{epsfig}
\newcommand{\Fig}[1]{Fig.~\ref{#1}}

\newcommand{\hmpc}{\ensuremath{\,h^{-1}\,{\rm Mpc}\,}}
\newcommand{\hgpc}{\ensuremath{\,h^{-1}\,{\rm Gpc}\,}}
\newcommand{\mpch}{\, \text{Mpc}/h}

\newcommand{\avg}[1]{\ensuremath{\langle #1 \rangle}}
\newcommand{\bma}{\begin{math}}
\newcommand{\ema}{\end{math}}
\newcommand{\beq}{\begin{equation}}
\newcommand{\eeq}{\end{equation}}
\newcommand{\beqa}{\begin{eqnarray}}
\newcommand{\eeqa}{\end{eqnarray}}
\newcommand{\bc}{\begin{center}}
\newcommand{\ec}{\end{center}} 
\newcommand{\bit}{\begin{itemize}}
\newcommand{\eit}{\end{itemize}}

\newcommand{\axi}{\left\langle x_i \right\rangle}
\newcommand{\rt}{R_{\text{T}}}
\font\BFd=cmmib10
\font\BFt=cmmib10
\font\BFs=cmmib10 scaled 700
\font\BFss=cmmib10 scaled 500

\def\bbox#1{%
\relax\ifmmode
\mathchoice
{{\hbox{\BFd #1}}}
{{\hbox{\BFt #1}}}
{{\hbox{\BFs #1}}}
{{\hbox{\BFss #1}}}
\else \mbox{#1} \fi }

\def\k{{\bbox{k}}}

\newcommand{\MHz}{\,\mbox{MHz}}
\newcommand{\meter}{\,\mbox{m}}
\newcommand{\kelvin}{\,\mbox{K}}
\newcommand{\xhi}{x_{\text{HI}}}
\newcommand{\axhi}{\langle x_{\text{HI}}\rangle}
\newcommand{\dd}{\text{d}}

\begin{document}

\submitted{\today. To be submitted to \apj.} 

\title{Identifying Ionized Regions in Noisy Redshifted 21 cm Data Sets}
\author{Matthew Malloy\altaffilmark{1} \& Adam Lidz\altaffilmark{1}}
\altaffiltext{1} {Department of Physics \& Astronomy, University of Pennsylvania, 209 South 33rd Street, Philadelphia, PA 19104, USA}
\email{mattma@sas.upenn.edu}

\begin{abstract}
One of the most promising approaches for studying reionization is to
use the redshifted 21 cm line. Early generations of redshifted 21 cm
surveys will not, however, have the sensitivity to make detailed maps
of the reionization process, and will instead focus on statistical
measurements. Here we show that it may nonetheless be possible to {\em directly identify ionized
  regions} in upcoming data sets by applying suitable filters to the noisy data. The locations
of prominent minima in the filtered data correspond well with the positions of ionized regions.
In particular,
we corrupt semi-numeric simulations of the redshifted 21 cm signal
during reionization with thermal noise at the level expected for a 500
antenna tile version of the Murchison Widefield Array (MWA), and mimic
the degrading effects of foreground cleaning. Using a matched filter
technique, we find that the MWA should be able to directly identify
ionized regions despite the large thermal noise. In a plausible fiducial 
model in which $\sim 20\%$ of
the volume of the Universe is neutral at $z \sim 7$, we find that a
500-tile MWA may directly identify as many as $\sim 150$ ionized
regions in a $6$ MHz portion of its survey volume and roughly determine the size of each of
these regions. This may, in turn,
allow interesting multi-wavelength follow-up observations, comparing
galaxy properties inside and outside of ionized regions. We discuss
how the optimal configuration of radio antenna tiles for detecting
ionized regions with a matched filter technique differs
from the optimal design for measuring power spectra. These considerations
have potentially important implications for the design of future
redshifted 21 cm surveys.
\end{abstract}


\section{Introduction} \label{sec:intro}
 
The Epoch of Reionization (EoR) is the time period when early generations
of galaxies first turn on and gradually photoionize neutral hydrogen
gas in the surrounding intergalactic medium (IGM). The IGM
is expected to resemble a two phase medium during reionization.
One phase consists of highly ionized regions, termed `ionized bubbles', that form around clustered
groups of ionizing sources, while the other phase is made up of intervening mostly neutral regions that
shrink and eventually vanish as reionization progresses. A primary goal
of reionization studies is to determine the size distribution and volume-filling factor of
the ionized bubbles. This should, in turn, significantly improve our understanding of high
redshift galaxy and structure formation. 
A wide variety of current
observations have started to provide tantalizing hints regarding the timing and nature of the EoR (e.g., \citealt{Fan:2005es}, \citealt{Totani:2005ng}, 
\citealt{Dunkley:2008ie}, \citealt{Ouchi:2010}, \citealt{Bouwens:2011xu}, \citealt{Mortlock:2011va}, \citealt{Zahn:2011vp}, \citealt{Schenker:2011ea}),
but we still await a more detailed understanding. 

A highly anticipated way of improving our knowledge of the EoR is to directly detect intergalactic neutral hydrogen
from the EoR using the redshifted 21 cm transition (e.g., \citealt{Madau:1996cs}, \citealt{Zaldarriaga:2003du}, \citealt{Furlanetto:2006jb}). Indeed, several radio telescopes have been constructed, or are presently under construction, in effort to detect this signal, including the Giant Metrewave Radio Telescope (GMRT) (\citealt{Paciga:2010yy}), the Low Frequency Array (LOFAR) (\citealt{Harker:2010ht}), the Murchison Widefield Array (MWA) (\citealt{Lonsdale:2009cb}), and the Precision Array for Probing the Epoch of Reionization (PAPER) 
(\citealt{Parsons:2009in}). This method provides the most direct, and potentially most
powerful, way of studying reionization, but several challenges need first to be overcome.
In particular, upcoming surveys will need to extract the faint cosmological signal in the presence of strong
foreground emission from our own galaxy and extragalactic
point sources, and to control systematic effects from the instrumental response, polarization leakage, calibration errors,
and other sources of contamination (e.g., \citealt{Liu:2009qga}, \citealt{Datta:2010pk}, \citealt{Harker:2010ht}, \citealt{Petrovic:2010me}, \citealt{Morales:2012kf}, \citealt{Parsons:2012qh}). In addition, thermal noise will prevent early generations of 21 cm experiments from making detailed maps
of the reionization process. Instead, detections will mostly be of a statistical
nature (\citealt{McQuinn:2005hk}). For example, a primary goal of these 
experiments is to measure the power spectrum of 21 cm brightness
temperature fluctuations by binning together many individually noisy Fourier 
modes (\citealt{Zaldarriaga:2003du}, \citealt{Morales:2003vn}, \citealt{Bowman:2005cr},
\citealt{McQuinn:2005hk}). 

It is unclear, however, how best to analyze the upcoming redshifted 21 cm data. Most previous work
has focused only on the power spectrum of 21 cm brightness temperature fluctuations (e.g., \citealt{Furlanetto:2004ha}, \citealt{Lidz:2007az}, \citealt{Mesinger:2010ne}). This statistic does not provide a complete
description of the 21 cm signal from the EoR, which will be highly non-Gaussian, with large ionized regions of essentially zero signal
intermixed with surrounding neutral regions. The power spectrum, and especially its redshift evolution, do encode interesting information
about the volume-averaged ionized fraction and the bubble size distribution (e.g., \citealt{Lidz:2007az}). However, these inferences are somewhat indirect and likely model dependent,
and so it is natural to ask if there are more direct ways of determining the properties of the ionized regions.

The approach we explore here is to check whether it may be possible to directly identify ionized regions in
noisy redshifted 21 cm observations by applying suitable filters to the noisy data. Our aim here is to blindly identify ionized bubbles across an entire survey volume, rather than to consider targeted searches around special
regions, such as those containing known quasars (e.g., \citealt{Wyithe:2004ta}, \citealt{Friedrich:2012fy}).
Since the 21 cm signal from reionization is
expected to have structure on rather large scales -- $\gtrsim 30$ $h^{-1}$Mpc co-moving (\citealt{Furlanetto:2004nh}, \citealt{Iliev:2005sz}, \citealt{Zahn:2006sg}, \citealt{McQuinn:2006et}) -- it may be possible to make
crude images of the large scale features even in the regime where the signal to noise per resolution element is much less
than unity.  Even if it is only possible to identify a few unusually large ionized regions in upcoming data sets, this would
still be quite valuable. Any such detection would be straightforward to interpret, and would open-up several interesting
possibilities for follow-up investigations.
Towards this end, we extend previous work by \cite{Datta:2007nj} and \cite{Datta:2008ry}, who considered the prospects for detecting
ionized regions using an optimal matched filter. A matched filter is constructed by correlating a known `template' signal with a
noisy data set in order to determine whether the template signal is present in the noisy data. Matched filters are used
widely in astrophysics: to name just a few examples, matched filters are used to detect clusters in cosmic microwave background (CMB) data (\citealt{Haehnelt:1995dg}), to identify galaxy clusters from weak lensing shear fields (e.g., \citealt{Hennawi:2004ai}, \citealt{Marian:2008fd}), 
and are central to data analysis efforts aimed at detecting gravitational waves (e.g., \citealt{Owen:1998dk}).

The outline of this paper is as follows. In \S\ref{sec:method} we describe the mock 21 cm data sets used in our investigations.
We use the mock data to first consider the ability of future surveys
to make maps of the redshifted 21 cm signal (\S\ref{sec:imaging}). In \S\ref{sec:ionprospects}, 
we then quantify the prospects for identifying individual ionized regions using
a matched filter technique. In \S\ref{sec:Variations} and \S\ref{sec:LOFAR} we
consider variations around our fiducial choice of reionization history and redshifted
21 cm survey parameters. We compare with previous related work in \S\ref{sec:PreviousWork},
and conclude in \S \ref{sec:Conclusion}.
Throughout we consider a $\Lambda$CDM cosmology
parametrized by $n_s =1, \sigma_8 = 0.8, \Omega_m = 0.27,
\Omega_\Lambda = 0.73, \Omega_b = 0.046$, and $h=0.7$, (all symbols
have their usual meanings), consistent with the latest WMAP
constraints from \cite{Komatsu:2010fb}.

\section{Method} \label{sec:method}
 
Briefly, our approach is to construct mock redshifted 21 cm data sets and check whether we
can successfully identify known `input' 
ionized regions in the presence of realistic levels of instrumental noise and the degrading impact
of foreground cleaning. Here we describe the ingredients of our mock data sets: our
simulations of reionization and the 21 cm signal, our model for thermal noise, and
our approach for incorporating the impact of foreground cleaning.

\subsection{The 21 cm Signal} \label{sec:21cm}

First, let us describe the underlying 21 cm signal and our reionization simulations.
The 21 cm signal will be measured through its contrast with the cosmic microwave background (CMB). The brightness temperature contrast
between the CMB and the 21 cm line from a neutral hydrogen cloud with neutral fraction $\xhi$ and fractional baryon overdensity
$\delta_\rho$ is (\citealt{Zaldarriaga:2003du}):
\begin{equation}
\delta T_{\text{b}} = 28 \xhi (1 + \delta_{\rho})\left( \frac{T_{\text{S}} - T_{\gamma}}{T_{\text{S}}} \right) \left( \frac{1 + z}{10} \right)^{1/2}\text{mK}.
\end{equation}
Here $T_\gamma$ denotes the CMB temperature and $T_{\rm S}$ is the spin temperature of the 21 cm line. Here and throughout
we neglect effects from peculiar velocities, which should be a good approximation at the redshifts and neutral fractions
of interest (e.g., \citealt{Mesinger:2007pd}, \citealt{Mao:2011xp}). Furthermore, throughout we assume that the spin temperature is globally much larger than
the CMB temperature, i.e., we assume that
$T_{\text{S}} \gg T_{\gamma}$. In this case the 21 cm signal appears in emission and the brightness temperature contrast
is independent of $T_{\rm S}$. This is expected to be a good approximation for the volume-averaged ionized fractions of interest for our
present study, although it will break down at earlier times (e.g., \citealt{Ciardi:2003hg}). With these approximations, 
\begin{equation}
\delta T_{\text{b}} = T_{0} \xhi (1 + \delta_{\rho}), \label{eq:21cm}
\end{equation}
 where $T_{0} = 28\left[ (1+z)/10 \right]^{1/2}$mK. Throughout this paper, we refer to the brightness temperature contrast 
in units of $T_{0}$.

\subsection{Semi-Numeric Simulations} \label{sec:sims}

In order to simulate reionization we use the `semi-numeric' scheme described in 
\cite{Zahn:2006sg} (see also e.g., \citealt{Mesinger:2010ne}, for related work and extensions to this technique). This scheme is essentially a Monte Carlo implementation
of the analytic model of \cite{Furlanetto:2004nh}, which is in turn based
on the excursion set formalism. The \cite{Zahn:2006sg} algorithm allows us to rapidly generate
realizations of the ionization field over large simulation volumes at various
stages of the reionization process. The results of these calculations agree well with
more detailed simulations of reionization on large scales (\citealt{Zahn:2006sg,Zahn:2010yw}). 

We start by generating a realization of the linear density field in a simulation
box with a co-moving side length of $1$ $h^{-1}$Gpc and $512^3$ grid cells. The
ionization field, $x_i$, is generated following the algorithm of \cite{Zahn:2006sg},
assuming a minimum host halo mass of $M_{\rm min} = 10^8 M_\odot$, comparable
to the atomic cooling mass at these redshifts (\citealt{Barkana:2000fd}). Each halo above $M_{\rm min}$
is assumed to host an ionizing source, and the ionizing efficiency of each galaxy is taken to be 
independent of halo mass. In our fiducial model, we adjust the ionizing efficiency
so that the volume-averaged ionization fraction is $\avg{x_i} = 0.79$ at
$z_{\rm fid} = 6.9$. We focus most of our analysis on this redshift and on this
particular model for the volume-averaged ionized fraction. However, we consider additional 
redshifts in \S\ref{sec:Variations}, as well as variations around our fiducial
ionization history in effort to bracket current uncertainties in the ionization history (see 
e.g., \citealt{Kuhlen:2012vy}, \citealt{Zahn:2011vp}).

From the linear density field
and the ionization field we generate the 21 cm brightness temperature
contrast following Equation \ref{eq:21cm}.  Using the linear density field here -- rather
than the evolved non-linear density field -- should be a good approximation for the large
scales of interest for our study; we focus on length scales of $R \gtrsim 20$ $h^{-1}$Mpc and high redshift ($z \gtrsim 6$) 
in subsequent sections. 

\subsection{Redshifted 21 cm Surveys and Thermal Noise} \label{sec:noise}

We mostly consider two concrete examples of upcoming 
redshifted 21 cm surveys. The first is based on the current, 128-tile version of the
MWA (\citealt{Tingay:2012ps}) and the second is based on an expanded, 500-tile version
of the MWA (as described in
\citealt{Lonsdale:2009cb}, and considered in previous work such as \citealt{McQuinn:2005hk, Lidz:2007az}).
These two examples are intended to indicate the general prospects for imaging and bubble identification with
first and second generation 21 cm surveys, respectively. Similar considerations would apply
for other experiments, but we choose these as a concrete set of examples.
We mainly focus on the 500-tile configuration in this paper because of its greater sensitivity. In 
\S\ref{sec:128tiles}, we shift to considering 128-tile configurations
and in \S\ref{sec:LOFAR}, we consider a LOFAR-style interferometer for comparison. Hereafter, we refer
to the 500-tile configuration as the MWA-500 and the 128-tile version as the MWA-128.

Throughout this paper, we work in co-moving coordinates described by Cartesian
labels ($x$-$y$-$z$), with Fourier counterparts ($k_x$-$k_y$-$k_z$). The Fourier
modes can be connected directly with the $u$-$v$-$\nu$ coordinate system generally
used to describe interferometric measurements. Here $u$ and
$v$ describe the physical separation of a pair of antennae in units of
the observed wavelength, while $\nu$ describes the corresponding
observed frequency. The instrument makes measurements for every frequency, $\nu$, in
its bandwidth, and for every antenna tile separation, $(u,v)$, sampled by the
array. In order to shift to a
Fourier space description, the interferometric measurements must first be Fourier-transformed
along the frequency direction. 
With our Fourier convention, the relation between the two sets of coordinates
is given by:
\begin{equation}
k_{x} = \frac{2\pi u}{D} \hspace{1cm} k_{y} = \frac{2\pi
  v}{D} \hspace{1cm} k_z = \frac{2\pi}{\Delta \chi}, 
\label{eq:UtoK}\end{equation}
where $D$ is the co-moving distance to the survey center and $\Delta
\chi$ is the co-moving distance corresponding to a small difference in observed
frequency of $\Delta \nu$ (e.g., \citealt{Liu:2009qga}). For small $\Delta \nu/\nu$, we can express $\Delta \chi$ as
\begin{equation}
\Delta \chi \approx \frac{c (1 + z_{\rm fid})}{H(z_{\text{fid}})} \frac{|\Delta
  \nu|}{\nu},
\end{equation}
where $H(z_{\rm fid})$ is the Hubble parameter at the fiducial
redshift, and $|\Delta \nu|/\nu$ is the absolute value of the fractional difference between two nearby observed frequencies.

In order to test the prospects for imaging and bubble identification with
the MWA, we must corrupt the underlying 21 cm signal described in \S \ref{sec:sims}
with thermal noise. We do this by generating a Gaussian random noise field in
the $\k$-space coordinate system described above, using an appropriate power spectrum. We assume that
the co-variance matrix of the thermal noise power is diagonal in $\k$-space. We add
the resulting noise field to the underlying 21 cm signal (Equation \ref{eq:21cm}).
The power spectrum of the thermal noise is given by (\citealt{McQuinn:2005hk}, \citealt{Furlanetto:2006pg}):
\begin{equation}
P_{\text{N}}(k,\mu) =
\frac{T_{\text{sys}}^2}{Bt_{\text{int}}}\frac{D^2\Delta
  D}{n(k_\perp)}\left( \frac{\lambda^2}{A_{\text{e}}}
\right)^{2}. \label{eq:NoisePower}
\end{equation}
Here $\mu$ is the cosine of the angle between wavevector $k = |{\bf
  k}|$ and the line of sight, so that $k_{\perp} = \sqrt{1-\mu^2} k$ is the transverse component of the wavevector. We 
assume a system temperature of $T_{\text{sky}} = 280
\left[(1+z)/7.5\right]^{2.3}\kelvin$ (\citealt{Wyithe:2007if}) and a total
observing time of $t_{\text{int}} = 1000\ \text{hours}$, which is an optimistic
estimate for the observing time in one year. At our fiducial redshift of $z_{\rm fid} = 6.9$, the co-moving
distance to the
center of the survey is $D = 6.42 \times 10^3$ $h^{-1}$Mpc. In this equation, $\lambda$ denotes
the observed wavelength of the redshifted 21 cm line,
$\lambda = 0.211(1+z) \meter$, and $A_{\text{e}}$ is the effective area of
each antenna tile. We determine $A_{\text{e}}$ by linearly extrapolating
or interpolating from the values given in
\href{http://iopscience.iop.org/0004-637X/661/1/1/fulltext/64283.tb2.html}{Table
  2} of \cite{Bowman:2005hj}; the effective area at $z_{\rm fid} = 6.9$ is
$A_{\text{e}}=11.25 \meter^2$. We assume that the full survey bandwidth of
$32 \MHz$ is broken into individual blocks of bandwidth $B = 6 \MHz$ to protect
against redshift evolution across the analysis bandwidth (\citealt{McQuinn:2005hk}). The co-moving survey depth depth corresponding to a $B= 6 \MHz$ chunk is $\Delta D = 69$ $h^{-1}$Mpc.
The $n(k_\perp)$ term describes the configuration of the antenna tiles. More specifically,
it is the number density of baselines observing modes with transverse wavenumber
$k_{\perp}$ (\citealt{McQuinn:2005hk}). Following \cite{Bowman:2005cr} and \cite{McQuinn:2005hk},
we assume the antenna tiles are initially packed as closely as possible in a
dense compact core, and that the number density of antenna tiles subsequently falls off
as $r^{-2}$ out to a
maximum baseline of $1.5$ km. The radius of the dense antenna core
is set  by the requirement that the antenna density falls off as
$r^{-2}$ outside of the core, and that it integrates to the total
number of antennae. For the MWA-500, this gives $r_{\text{c}}
= 20 \meter$, while for the MWA-128, the core radius is $r_{\text{c}}
\approx 8 \meter$. Equation \ref{eq:NoisePower} gives the noise power spectrum in
units of mK$^2$, and so we divide by $T_0^2$ to combine with the simulated 21 cm signal expressed
in units of $T_0$.

Note that the volume of the MWA survey differs somewhat from that
of our reionization simulation. In particular, the transverse dimension
of the simulation is smaller than that of the MWA by a factor of $\sim 3$, while the simulation
is deeper in the line-of-sight direction by about the same factor, as compared with the full MWA bandwidth. However, we remove
the long wavelength modes along the line-of-sight direction to mimic foreground
cleaning (\S \ref{sec:foregrounds}), and so we do not, in practice, use the longer line-of-sight scales in our simulation
box. As we will see, the ionized regions in the simulation are substantially
smaller than the transverse length of the box. Transverse slices should therefore
be representative of what the actual MWA will observe from a fraction of its 
larger
field of view. We have checked that the coarser transverse $k$-space sampling in the simulation
compared to in the actual MWA survey does not impact our results.

\subsection{Foregrounds} \label{sec:foregrounds}

Next, we need to consider contamination from foreground emission 
at the frequencies of interest. The relevant foregrounds include
diffuse Galactic synchrotron radiation, extragalactic point
sources, and Galactic Bremsstrahlung radiation. While these
foregrounds are many orders of magnitude brighter than the
cosmological 21 cm signal, they are expected to individually follow
smooth power laws in frequency. Over a sufficiently small frequency range,
the summed contributions can also be approximated as
following a smooth power law, while the 21 cm signal will vary
rapidly. This allows the foregrounds to be removed from the data
by, for example, fitting a low-order function along each line of sight and subtracting
it. While this procedure is effective at removing foreground contamination, it also removes long wavelength
modes along the line of sight from the signal itself, and hence prevents measuring these modes. 
Several related methods for foreground removal have been discussed in the
literature (e.g., \citealt{Wang:2005zj},
\citealt{Harker:2009hg}, \citealt{Petrovic:2010me},
\citealt{Chapman:2012yj}). In this work, we approximately mimic the
degrading effects from foreground removal by subtracting the running
mean from the noisy signal along each line of sight, rather than
including realizations of the foregrounds in our simulation and excising
them with one of the above algorithms.
We generally remove the running mean over a bandwidth of $16\MHz$, which corresponds
to a co-moving distance of $L_{\text{fg}} = 185\hmpc$ at redshift $z_{\rm fid} = 6.9$; we consider the impact of other choices of $L_{\text{fg}}$ in \S \ref{sec:ForegroundCleaning}.
We defer more detailed models of foreground contamination, and foreground
removal algorithms, to future work.

\section{Prospects for Imaging} \label{sec:imaging}
 
Having described our mock 21 cm data sets, we now turn to
consider the prospects for constructing direct `images' of
the redshifted 21 cm signal. Previous studies already suggest
that the prospects for imaging with the
MWA-500 are limited (\citealt{McQuinn:2005hk}). Here we emphasize that
even a crude, low-resolution image of the redshifted 21 cm signal may be
quite interesting, especially given that the ionized regions during reionization may be rather large scale
features. We hence seek to quantify the imaging capabilities
further using our corrupted reionization
simulations. Here our work complements recent work in a similar
vein by \cite{Zaroubi:2012cy}, who considered the prospects for imaging with
LOFAR. While the central idea in this section
is similar to this previous work, we focus on the MWA while \cite{Zaroubi:2012cy}
considered LOFAR. In order to construct the best possible images from
the noisy mock 21 cm data, we apply a Wiener filter. We assess the
ability of the MWA to image the redshifted 21 cm sky by comparing the filtered (recovered)  noisy
signal with the underlying noise-free 21 cm input signal.

\subsection{The Wiener Filter} \label{sec:wiener}
 
The Wiener filter is the optimal filter for extracting
an input signal of known power spectrum when it is corrupted by additive noise,
also with known power spectrum. 
As described in \cite{NRecipes},
this filter is optimal in that it minimizes the expectation value of the integrated squared
error between the estimated signal field and the true signal
field. The estimate of the true signal is a convolution of the Wiener
filter and the corrupted signal in real space, and so is a product of
the two quantities in Fourier space,
\begin{equation}
\tilde{S}({\bf k}) = C({\bf k})W({\bf
  k}), \label{eq:FourierConvolution}
\end{equation}
where $C({\bf k})$, $W({\bf k})$, and $\tilde{S}({\bf k})$ are the
Fourier transforms of the corrupted signal, Wiener filter, and
estimated signal, respectively. Requiring that the filter be optimal
in the least-square sense results in $W({\bf k})$ taking the form
\begin{equation}  W(k,\mu) = \frac{P_{\text{S}}(k)}{P_{\text{S}}(k) + P_{\text{N}}(k,\mu)}, \label{eq:WienerFilter} \end{equation}
where $P_{\text{S}}(k)$ and $P_{\text{N}}(k,\mu)$ are the power
spectra of the signal and noise, respectively. We note that, while the
signal power spectrum is roughly isotropic\footnote{Redshift-space distortions
and redshift evolution across the observed bandwidth break isotropy (e.g. \citealt{Datta:2011hv}). However,
for the bandwidth considered here ($B = 6 \MHz$) and the neutral fractions of interest, the
signal should be approximately isotropic.}, the noise power spectrum depends on $\mu$ and
consequently so does the filter. The filter keeps a unity weighting
for $k$-modes where $P_{\text{S}}({\bf k}) \gg P_{\text{N}}({\bf k})$
and significantly downweights $k$-modes where $P_{\text{S}}({\bf k})
\ll P_{\text{N}}({\bf k})$. This can allow for partial recovery of
the original signal, provided that the signal power dominates for some
$\k$-modes.

The Wiener filter requires an estimate of the signal power spectrum, $P_{\text{S}}(k)$, and of
the total (signal plus noise) power spectrum, $P_{\text{S}}(k) + P_{\text{N}}(k, \mu)$, as inputs. These
may not be precisely known. However, since the filter is the outcome of a
minimization problem -- i.e., it minimizes the expected difference
between the estimated and true fields -- the accuracy of the filter
should be insensitive to small changes about its optimal value. In
other words, the accuracy of the filter is not expected to
change greatly by using estimates of the signal and noise power spectra rather than
the true spectra. 

Furthermore, we do expect to have an estimate of the underlying signal power
spectrum; measuring this statistic is a major goal of redshifted 21 cm surveys. Specifically, the underlying
signal power can be estimated by cross-correlating
redshifted 21 cm measurements made over 
two different time intervals (after foregrounds have been removed). The statistical
properties of the signal should be identical across the two different time
periods, but the thermal noise contributions will be independent. The
cross-correlation between two time chunks then provides an unbiased
estimate of the signal power (e.g., \citealt{Liu:2009qga}). Estimates of the noise power spectrum can then be made by
subtracting the estimated signal power from the power measured over
the entire integration time, which contains both
the signal and noise contributions. The Wiener filter does not
actually require the noise power spectrum to be known on its
own. However, in  \S\ref{sec:ionprospects} we consider the optimal
matched filter, which does have this requirement. Throughout this study, we
assume perfect knowledge of the underlying power spectra.

Before applying the Wiener filter to our corrupted simulations, it is
useful to estimate the expected signal-to-noise ratio of the filtered maps 
analytically, using
simulated signal power spectra and the noise power
spectrum of Equation \ref{eq:NoisePower}. The expected signal-to-noise ratio of
the Wiener-filtered field is $\mathcal{S}_{\text{wf}} =
\tilde{\sigma}_{\text{S}}/\tilde{\sigma}_{N}$, where
$\tilde{\sigma}^{2}_{\text{S(N)}}$ is the filtered signal (noise)
variance. The signal and noise variance can in turn be calculated as
integrals over their respective power spectra,
\begin{equation}
\tilde{\sigma}^2_{\text{S(N)}} = \int \frac{\dd^3
  k}{(2\pi)^3}|W(k,\mu)|^{2}P_{\text{S(N)}}(k,\mu). \label{eq:snr}
\end{equation}
Here we use $P_{\text{S(N)}}$ to denote the power spectrum of the
signal (noise). One can also consider the impact of foreground
cleaning here by downweighting modes where the foreground power
is large compared to the signal power.
In order to consider the dependence of the signal-to-noise ratio
on the stage of reionization, we consider simulation outputs
in which the volume-averaged ionization fraction is 
$\avg{x_i} = 0.51, 0.68, 0.79$ and $0.89$. We consider each of these
models at our fiducial redshift of $z_{\rm fid} = 6.9$.\footnote{In practice,
the simulated ionization fields for ionized fractions lower (higher) than our fiducial value ($\avg{x_i} = 0.79$ at $z_{\rm fid} = 6.9$) 
come from slightly higher
(lower) redshift simulation outputs. We generate the 21 cm signal and noise as though each data cube were in
fact at $z_{\rm fid} = 6.9$. This is appropriate to the extent that
the statistical properties of the ionized regions are mainly determined by
the volume-averaged ionized fraction, and are relatively insensitive to the precise
redshift at which a given ionized fraction is reached (see \citealt{McQuinn:2006et}
and \citealt{Furlanetto:2004nh}.)} Presently, we don't consider still earlier stages
of reionization since the prospects for imaging with the MWA-500 are especially poor for lower
ionized fractions.

\begin{figure}[h]
  \centering
  \includegraphics[width=9cm]{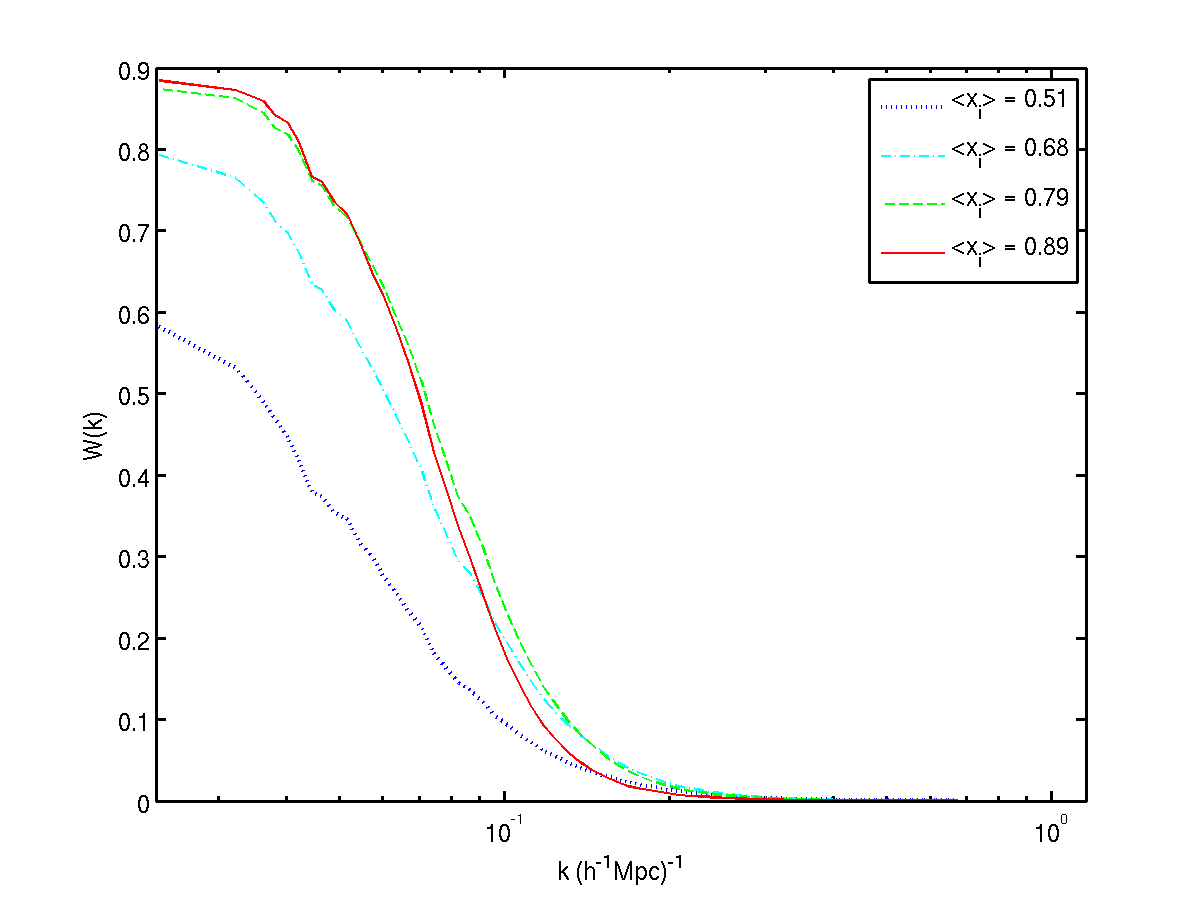}  
  \caption{Fourier profile of the Wiener filter, $W(k)$. The filter is averaged over
    line-of-sight angle and the results are shown at $z_{\rm fid} = 6.9$ for
simulated models with $\axi = 0.51$ (blue dotted), $\axi = 0.68$
    (cyan dot-dashed), $\axi = 0.79$ (green dashed), and $\axi = 0.89$ (red solid).}
  \label{fig:WienerFilter}
\end{figure}

The resulting Wiener filters for the different values of $\avg{x_i}$
are shown in Figure \ref{fig:WienerFilter}, after integrating over
angle $\mu$. In this figure, foreground cleaning has been accounted for by subtracting a running mean along the line of sight, as described in \S\ref{sec:foregrounds}. It is helpful to note, from Equation
\ref{eq:WienerFilter}, that the filter is equal to $1/2$ for modes
where the signal and noise power are equal. The figure suggests that a
small range of $k$-modes with $k \lesssim 0.1 h$ Mpc$^{-1}$ will have
signal-to-noise ratio larger than unity for all four ionized fractions
considered, although imaging is less promising for the smaller ionized
fractions. If the ionized regions are larger than in our fiducial
model --  as expected if, for example, rarer yet more efficient and
more clustered sources dominate reionization
(e.g., \citealt{McQuinn:2006et}, \citealt{Lidz:2007az}) -- then the
prospects for imaging may improve somewhat. Performing the integrals
in Equation \ref{eq:snr}, while incorporating foreground cleaning, we
find that the total signal-to-noise ratio expected for the MWA-500 is
$\mathcal{S}_{\text{wf}} = 0.52$, $0.79$, $1$, and $1.2$ for
$\avg{x_i} = 0.51, 0.68, 0.79, 0.89$, respectively.

\subsection{Application to a Simulated 21 cm Signal} \label{sec:wapplied}

\begin{figure}[t]
  \includegraphics[width=8.4cm]{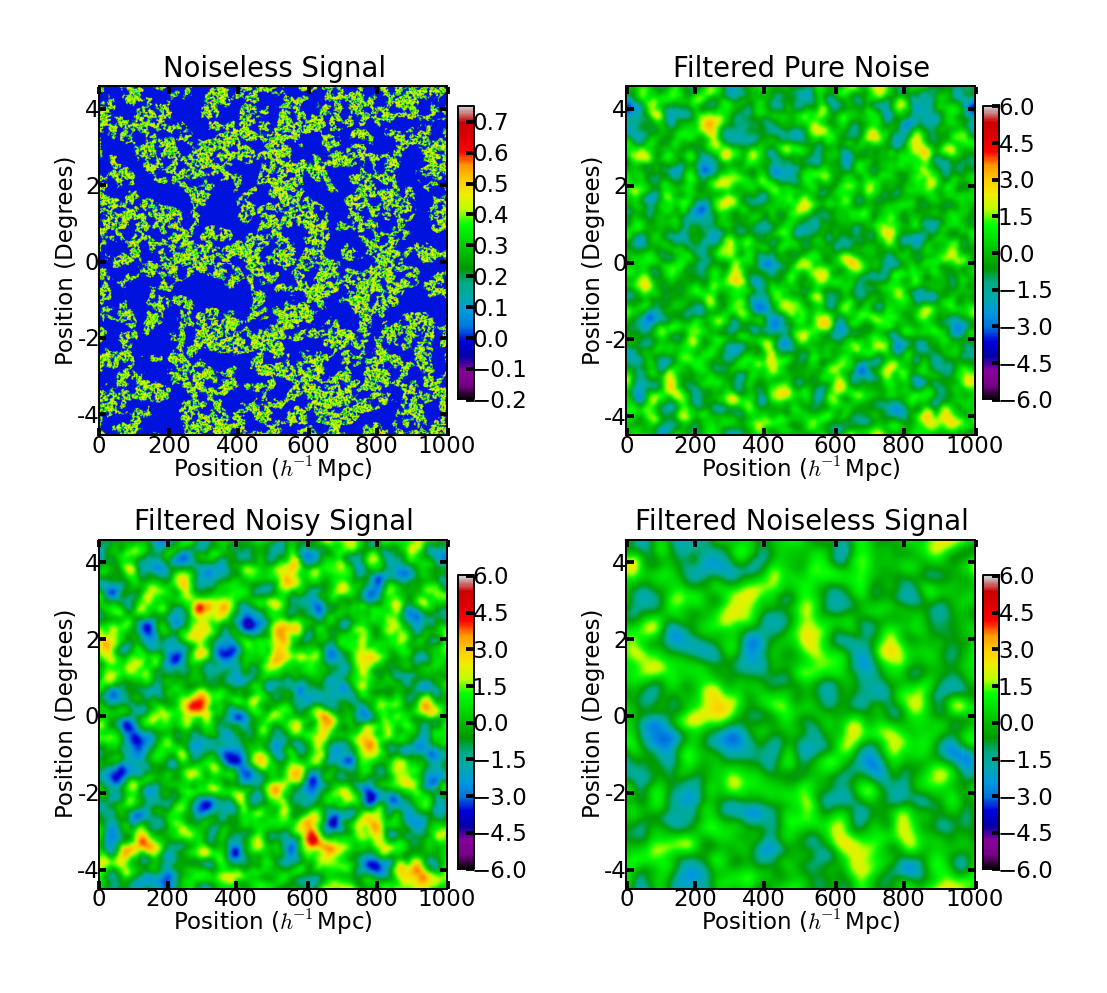}
  \caption{Application of the Wiener filter to simulated data. The
results are for our fiducial model with $\avg{x_i} = 0.79$ at $z_{\rm fid} = 6.9$. \textit{Top-Left}:
    Spatial slice of the unfiltered and noise-less 21 cm brightness
    temperature contrast field (normalized by
    $T_0$). \textit{Top-Right}: Simulated signal-to-noise field after
    applying the Wiener filter to a pure noise
    field. \textit{Bottom-Left}: Simulated signal-to-noise field
    after applying the Wiener filter to the noisy
    signal. This can be compared with the uncorrupted input signal shown
in the top-left panel and the noise realization in the top-right panel. \textit{Bottom-Right}: Simulated signal-to-noise field after applying the Wiener filter to the noiseless signal. (The filtered noiseless signal shown here is normalized by the standard deviation of the noise to facilitate comparison with the other panels.) All panels show a square section of the MWA field of
    view transverse to the line of sight with sidelength $L = 1 \hgpc$. All slice thicknesses are $\sim 8 \hmpc$. Unless noted otherwise, the simulation slices in subsequent figures have these same dimensions.}
  \label{fig:Wiener}
\end{figure}

\begin{figure}
  \centering
  \includegraphics[width=9cm]{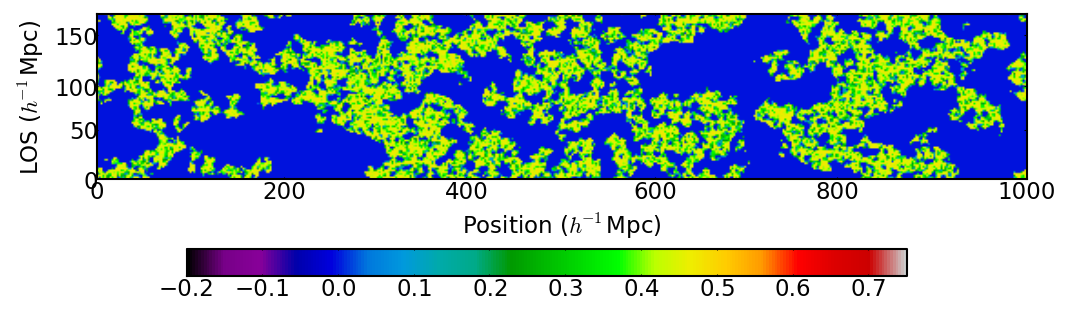}
  \includegraphics[width=9cm]{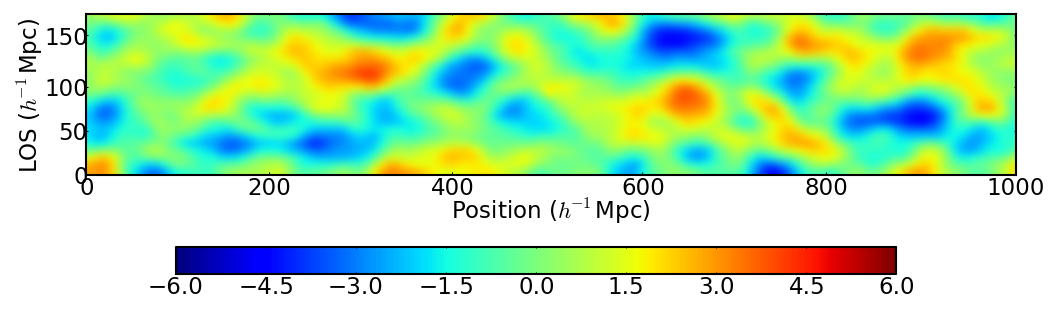}
  \caption{Impact of foreground cleaning on the Wiener-filtered field. The top slice is a perpendicular, zoomed-in view of the simulated, unfiltered, noise-less brightness temperature contrast. 
The bottom slice is the signal-to-noise of the same region after applying the
    Wiener filter to the noisy signal field. The vertical axis shows the
    line-of-sight direction, with its extent set to the
    distance scale for foreground removal, $L_{\text{fg}} = 185\hmpc$. The horizontal axis shows a dimension transverse to the line of sight and extends 1$\hgpc$. }
  \label{fig:WienerLOS}
\end{figure}

With the analytic signal-to-noise ratio estimates as a guide, we apply
the Wiener filter to our mock noisy redshifted 21 cm data. The results
of these calculations, for a particular slice through the simulation
volume, are shown in Figure \ref{fig:Wiener}.  The side length ($1$ $h^{-1}$Gpc) of each slice
is a factor of $\sim 3$ smaller than the transverse dimension of the MWA. One can asses how well
the original signal is `recovered' by comparing the top-left panel of
the figure which shows the input signal with the bottom-left panel which
shows the filtered noisy signal, after mimicking foreground
removal. The two panels do not bear a striking resemblance since the
average signal-to-noise ratio is only of order
unity. Nonetheless, it is encouraging that many of the minima in 
the filtered noisy
signal do indeed correspond to ionized regions in the input signal.  
Furthermore, we can compare the filtered noisy signal in the
bottom-left panel with the top-right panel, which shows filtered pure
noise. While these two panels do not look drastically different, they
are easily distinguishable from each other given the increased
contrast in the filtered noisy signal. Note that the noisy signal -- in the absence of any filtering (not shown) -- looks like pure noise since the
signal to noise per resolution element is extremely low, and so filtering
helps significantly.
In addition, we see that the
filtered noisy signal obtains signal-to-noise values exceeding
$6-\sigma$, while the statistical significance of the filtered noise
does not exceed $\sim 5-\sigma$. Quantitatively, $\sim 3\%$
($\sim 0.03\%$) of the volume in the filtered noisy signal is occupied
by pixels with statistical significance greater (in magnitude) than
$3-\sigma$ ($5-\sigma$). This is expected given that the filtered data cube
has an average signal-to-noise ratio of $\sigma_{\text{S}}/\sigma_{\text{N}} \approx 1$, as anticipated in the analytic calculation
of \S\ref{sec:wiener}.

Comparing the filtered noisy signal and the filtered pure noise,
one can see that ionized regions in the underlying signal are
diminished if they happen to be coincident with upward fluctuations 
in the noise, as expected. For example, the ionized region in the
bottom-right corner of the unfiltered signal lies very close to a
$\sim 3-\sigma$ upward fluctuation in the filtered noise and, as a
result, appears with weak statistical significance in the filtered
noisy signal. Conversely, some of the most statistically
significant regions in the filtered noisy signal occur when ionized
regions overlap downward noise fluctuations. 
We can further compare the filtered noisy signal with the filtered noise-{\em less} signal,
shown in the bottom right panel of Figure \ref{fig:Wiener}. The filtered noise-less signal
is normalized by the standard deviation of the filtered noise so that it can be compared
with the signal-to-noise slices in the other panels.
This comparison reveals that high significance regions ($\gtrsim 5-\sigma$) in the 
filtered noisy signal only
line up well with the corresponding regions in the filtered noiseless signal if they
are coincident with downward fluctuations in the noise.
On its own, the filtered noiseless signal only attains statistical significances of $\lesssim4\sigma_{\text{N}}$.
Finally, Figure \ref{fig:WienerLOS} illustrates the impact of
foreground cleaning, performed here over a bandwidth of $16 \MHz$ (\S \ref{sec:foregrounds}). 
Foreground cleaning removes the long wavelength
modes along the line of sight -- which is along the vertical axis
in the figure -- and thereby compresses structures along the line of sight.
However, the cleaning process only impacts the long wavelength line-of-sight
modes which still leaves room to image other modes robustly.

Note that the slice thickness ($8$ $h^{-1}$Mpc) in Figure \ref{fig:Wiener} and \ref{fig:WienerLOS} is somewhat
arbitrary. However, the Wiener filter smooths out structure on significantly larger
scales than this (Figure \ref{fig:WienerFilter}), and so we expect
similar results for other values of the slice thickness, provided the
slice is thin compared to the cut-off scale of the filter. 
In practice, of course,
one can make many independent maps similar to Figure \ref{fig:Wiener} from
the MWA-500 or similar surveys. Collectively, our results mostly confirm
previous wisdom; the prospects for imaging with the MWA-500 are limited.
Nonetheless, it appears that a signal-to-noise ratio of order unity
is achievable,
suggesting that the MWA-500 {\em can} make low resolution maps of the reionization
process.

\section{Prospects for Identifying Ionized Regions} \label{sec:ionprospects}

We now shift our focus to discuss whether it may also be possible
to identify interesting individual features in upcoming 21 cm data cubes.
In particular, we aim to identify ionized regions in noisy 21 cm data sets
and, furthermore, to estimate the spatial center and approximate size of each ionized bubble.
For this purpose, we will use an optimal matched filter technique. As we
discuss, individual
ionized regions may be identifiable as {\em prominent minima in the
filtered field}.

\subsection{The Optimal Matched Filter} \label{sec:optimal}

The optimal matched filter is suited for the case of a corrupted 
signal containing a known feature that one would like to extract. The
filter acts in Fourier space by cross-correlating the corrupted signal
with a template describing the known feature, while 
downweighting $\k$-modes in the
corrupted signal by the noise power. The resulting form of the filter in
Fourier space,
$M(k, \mu)$, is 
\begin{equation} M(k,\mu) = \frac{T({\bf k})}{P_{\text{N}}(k,\mu)}, \label{eq:MatchedFilter} \end{equation}
where $T({\bf k})$ is the Fourier profile of the known feature. The
filter is optimal in the sense that it maximizes the signal-to-noise
ratio in the filtered data cube at the location of the feature being
extracted. While the Wiener filter requires an estimate of the signal and total (signal plus noise) power spectra,
the matched filter requires a good estimate of the template profile, $T(\k)$, and the noise power spectrum, $P_{N}(k,\mu)$. For our present application,
we would like templates describing the ionized regions.
An appropriate choice is not obvious; theoretical models predict that the
ionization state of the gas during reionization has a complex, and
somewhat uncertain, morphology, with ionized regions of a range of
sizes and shapes (\citealt{Iliev:2005sz}, \citealt{Zahn:2006sg},
\citealt{McQuinn:2006et}). However, we find that the simplest conceivable
choice of template filters, corresponding to completely ionized 
spherical bubbles of varying size, are nonetheless effective at
identifying ionized regions with a more realistic and complex
morphology. In this case, $T(\k;R)$ is just the Fourier transform
of a spherical top-hat of radius $R$ and is given by
\begin{equation} T({\bf k};\rt) = \frac{V}{k^3\rt^3}\left[ -k\rt\cos k\rt + \sin k\rt \right], \label{eq:Template} \end{equation}
with $V$ denoting the volume of the spherical top-hat. Note that the precise normalization of the filter is unimportant since
we are mainly interested in the signal-to-noise ratio here, in which case the overall normalization divides out.

\subsection{Application to Isolated Spherical Ionized Regions with Noise} \label{sec:toy}

It is instructive to first consider an idealized test case that can be
treated analytically before applying the matched filter to our
full mock 21 cm data sets. In particular, we consider the
case of an isolated, spherical, and highly ionized region placed at the
origin and embedded in
realistic noise. We assume that the neutral fraction exterior
to the ionized region is uniform, with a mass-weighted neutral fraction
of $\avg{x_{\rm HI} (1 + \delta_\rho)}$. Ignoring foreground contamination
for the moment, the 21 cm signal may be written as: 
\begin{equation}
\delta T_{\text{b}}({\bf x}) - \avg{\delta T_{\text{b}}} = \tilde{B}({\bf x};R_{\text{B}}) +
\tilde{N}({\bf x}),
\end{equation}
where $B({\bf x};R_{\text{B}})$ denotes our isolated bubble of radius
$R_{\text{B}}$,
and $N({\bf x})$ denotes the thermal noise contribution to the
signal. We have subtracted off the overall mean brightness temperature, $\avg{\delta T_{\text{b}}}$,
since this will not be measured in an interferometric observation. The tildes indicate
that the spatial average has been removed from each of the underlying signal and noise so
that $\tilde{B}({\bf x};R_{\text{B}})$ and $\tilde{N}({\bf x})$ each have zero mean. In
this case $\tilde{B}({\bf x};R_{\text{B}})$ has an inverted spherical top-hat profile,
\begin{equation}
\tilde{B}({\bf x};R_{\text{B}}) = \begin{cases} -\left\langle
  \xhi(1+\delta_{\rho}) \right\rangle   &\text{$|{\bf x}| <
    R_{\text{B}}$,} \\ 0  &\text{otherwise.}
\end{cases}
\label{eq:Btilde}
\end{equation}
The Fourier
transform of the isolated bubble is hence related to the Fourier transform of our
template by $\tilde{B}({\bf k};R_{\text{B}}) = -\left\langle
\xhi(1+\delta_\rho) \right\rangle T({\bf k};R_{\text{B}})$. Note that we express brightness temperatures
in units of $T_0$ (see Equation \ref{eq:21cm}), and so all quantities here are dimensionless.

It is straightforward to derive the expected signal-to-noise ratio at the center
of the isolated ionized region, and thereby gauge the prospects for bubble detection with
a matched filter technique. Let us assume that the radius, $R_{\rm B}$, of our template filter
is well matched to the true radius of the ionized region. This will maximize the expected
signal-to-noise ratio. Neglecting foregrounds for the moment, and using the fact that
the thermal noise has zero mean, we find that the signal-to-noise ratio at bubble center
for the optimal matched filter is: 
\begin{equation} \mathcal{S}(R_{\text{B}}) = - \left\langle \xhi (1 + \delta_\rho) \right\rangle \left[\int \frac{\dd^3 k}{(2\pi)^3}\frac{T^2(k;R_{\text{B}})}{P_{\text{N}}(k,\mu)}\right]^{1/2}. \label{eq:ToySNR} \end{equation}
For our sign convention, in which the template and ionized regions have opposite signs, this quantity is 
negative -- ionized bubbles are regions of low 21 cm signal. The contribution of a Fourier mode to the signal to noise
ratio
depends on the relative size of $T^2(k;R_{\text{B}})$ and $P_{\text{N}}(k,\mu)$: modes for which the
template is much larger than the noise power contribute appreciably to $\mathcal{S}(R_{\text B})$ while
modes dominated by the noise power are not useful. The signal-to-noise ratio depends on the
neutral fraction: a larger exterior neutral fraction increases the contrast between
an ionized bubble and the exterior, and hence boosts the detectability of the
ionized region. 
We would like to calculate the expected signal-to-noise ratio for ionized regions of different sizes and
for various volume-averaged
ionization fractions. To do this, we need to connect the volume-averaged ionized fraction with 
the mass-averaged fraction, $\avg{x_{\rm HI} (1 + \delta_\rho)}$, which enters into Equation \ref{eq:ToySNR}.
Here we should incorporate that large scale overdense regions are generally ionized before typical regions
during reionization, i.e., the neutral fraction and overdensity fields are anti-correlated. Defining $\delta_x = (x_{\rm HI} - \avg{x_{\rm HI}})/\avg{x_{\rm HI}}$,
we approximate $\avg{\delta_x \delta_\rho}$ as fixed at $\avg{\delta_x \delta_\rho} = -0.25$ throughout
the reionization process (\citealt{Lidz:2006vj}). 

\begin{figure}[h]
  \centering
  \includegraphics[width=9cm]{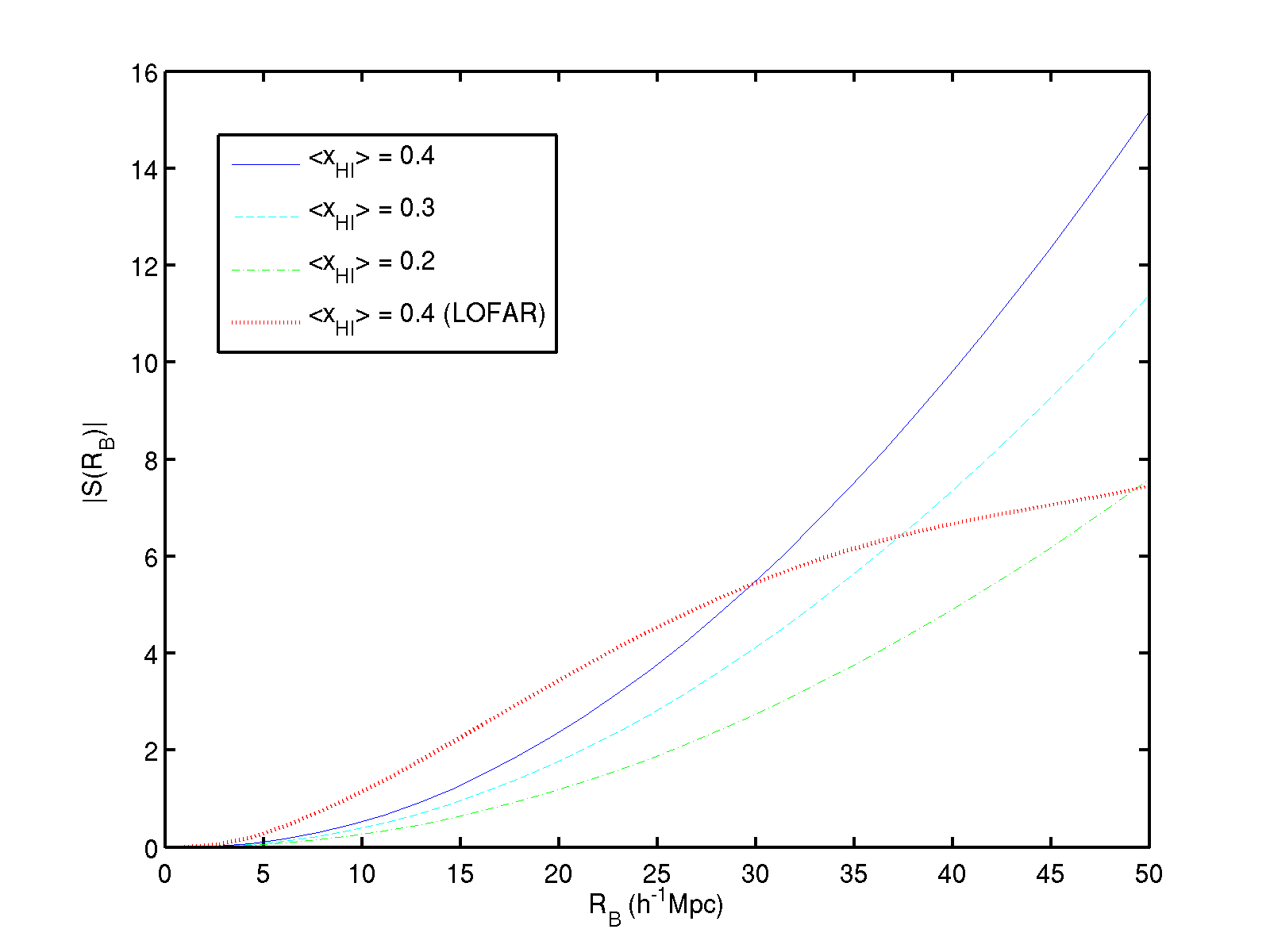}
  \caption{Expected signal-to-noise ratio at the center of
            isolated, spherical, ionized bubbles as a function of bubble
            radius after applying the optimal matched filter. The
curves show the signal-to-noise ratio at $z_{\rm fid} = 6.9$ for the
MWA-500 at various neutral fractions: $\axhi = 0.4$ (blue solid), $0.3$
            (cyan dashed), and $0.2$ (green dot-dashed). For contrast, the red dotted curve indicates the
            expected signal-to-noise for an interferometer 
with a field of view and collecting area
            similar to a 32-tile LOFAR-like antenna array (at $\axhi = 0.4$).}
  \label{fig:ToySNR}
\end{figure}

The results of the signal to noise calculation are shown in Figure
\ref{fig:ToySNR} for the MWA-500 and a LOFAR-type experiment. 
Here we consider only our
fiducial redshift, $z_{\rm fid} = 6.9$. The (absolute value of)
$\mathcal{S}(R_{\text{B}})$ is evidently a strongly increasing
function of bubble size. This occurs because the thermal noise is a
strong function of scale and only the rather large scale modes are
measurable. It is encouraging that the expected signal-to-noise ratio
exceeds five, $\mathcal{S}(R_{\text{B}}) \gtrsim 5$, for a range of
radii and neutral fractions. This corresponds to a $5-\sigma$
detection: `false' bubbles at this significance from downward
fluctuations in the noise are highly unlikely, with a fraction of 
only $\sim 3 \times 10^{-7}$ of pixels in the filtered noise having such a large
(negative) significance on their own. For simplicity we neglect
the impact of foreground cleaning in this figure: this will 
degrade the expected signal-to-noise ratios somewhat, as we will
consider subsequently (see \S \ref{sec:mapplied}, \S \ref{sec:ForegroundCleaning}).

In order to estimate the number of bubbles that can be detected from these curves, we need to consider
how many bubbles there are of different sizes, i.e., we need to fold in an estimate of the bubble
size distribution. In particular, while the contrast of an ionized region increases with the
neutral fraction, large ionized bubbles become increasingly scarce for larger values of the neutral
fraction. For instance, we can consider the model bubble size distributions in
\href{http://iopscience.iop.org/0004-637X/654/1/12/fulltext/65071.fg4.html}{Figure
  4} of \cite{Zahn:2006sg}. This figure indicates that bubbles of radius larger than $30$ $h^{-1}$Mpc
are exceedingly rare for neutral fractions larger than $\avg{x_{\rm HI}} > 0.5$, with only
the tail end of the distribution extending past $25$ $h^{-1}$Mpc. However, bubbles this size
are relatively common later in reionization. Since Figure \ref{fig:ToySNR} indicates
that only bubbles with $R \gtrsim 30$ $h^{-1}$Mpc exceed $\mathcal{S}(R_{\text{B}}) \gtrsim 5$,
this suggests that bubble detection is feasible for the MWA-500 after the Universe is more than
$\sim 50\%$ ionized, but that it will be difficult to use this method at earlier
stages of the reionization process. Also, note again that the calculation here neglects the effects of foreground cleaning. However, we find that incorporating foreground cleaning only has a small effect on bubbles of this size ($\lesssim 30\hmpc$, \S \ref{sec:ForegroundCleaning}). Bubble detection will also be challenging once
the Universe is less than $\sim 10-20\%$ neutral, owing mostly to the reduced contrast
between the bubbles and typical regions. If the ionized bubbles at a given stage of the EoR are larger than in
the model of \cite{Zahn:2006sg}, then the prospects for bubble detection will be enhanced. We refer
the reader to \cite{McQuinn:2006et} for a quantitative exploration of the bubble size distribution across
plausible models for the ionizing sources.

Finally, it is interesting to consider a LOFAR-style interferometer, as 
discussed further
in \S\ref{sec:LOFAR}. This is shown as the red dot-dashed curve in Figure \ref{fig:ToySNR}.
The expected $\mathcal{S}(R)$ exceeds that of the MWA-500 for small bubble radii, before
flattening off at larger radii. This occurs because the LOFAR-style interferometer
has more collecting area per baseline, but a larger minimum baseline. This makes it
more sensitive to the smaller ionized regions, but less sensitive to larger ones.

While the signal-to-noise curves in this toy case provide a useful guide, we should keep in mind their limitations. 
First, it considers only the case of a single isolated ionized region. Next, we consider here only
the signal to noise at the bubble center, while an ionized region will typically have
a strong (negative) signal to noise over much of its volume. Specifically, for bubbles with $10\mpch \lesssim R_{\text{B}} \lesssim 60\mpch$, we find that the signal-to-noise value at the edge of bubble is roughly half that at its center. This can help significantly with
detection. Finally, we consider only the {\em average} signal-to-noise ratio here. In
practice, the signal-to-noise ratio in a filtered map may fluctuate significantly around
this average, as we will see.

\subsection{Application to a Simulated 21 cm Signal} \label{sec:mapplied}

\begin{figure}[h]
  \centering
  \includegraphics[width=8.4cm]{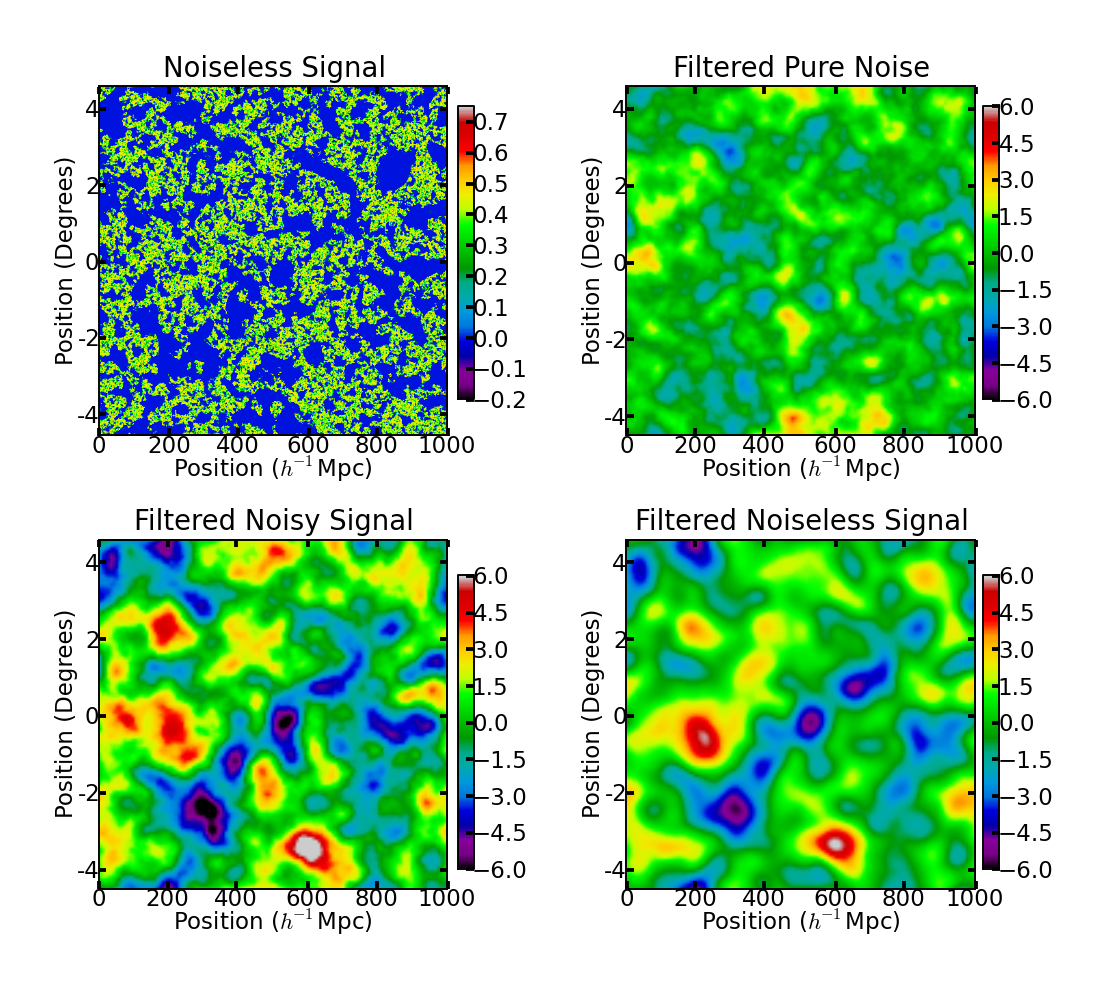}
  \caption{Application of the matched filter to simulated data and
    noise ($\langle x_i \rangle  = 0.79$ at $z_{\rm fid} = 6.9$). The 
    template radius
    of the filter is 35$\hmpc$, since this is a commonly detected
    bubble radius for our matched filter search. \textit{Top-Left}: Spatial slice of the
    unfiltered and noise-less 21 cm brightness temperature contrast
    field. \textit{Top-Right}: Simulated signal-to-noise field after
    applying the matched filter to a pure noise
    field. \textit{Bottom-Left}: Simulated signal-to-noise field
    after applying the matched filter to the noisy
    signal. This can be compared directly to the top-left panel. \textit{Bottom-Right}: Simulated signal-to-noise field after applying the matched filter to the noiseless signal. All panels are at the same spatial slice. See text for discussion on interpreting signal-to-noise values.}
  \label{fig:ThreeFigureMatched}
\end{figure}

\begin{figure}[h]
  \centering
  \includegraphics[width=9cm]{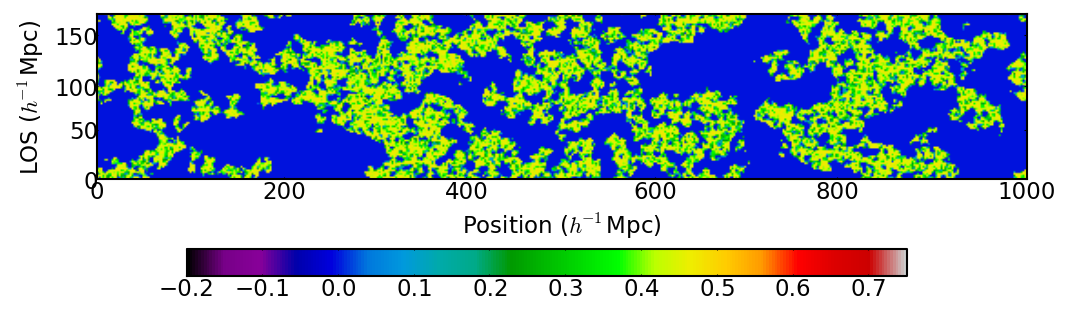}
  \includegraphics[width=9cm]{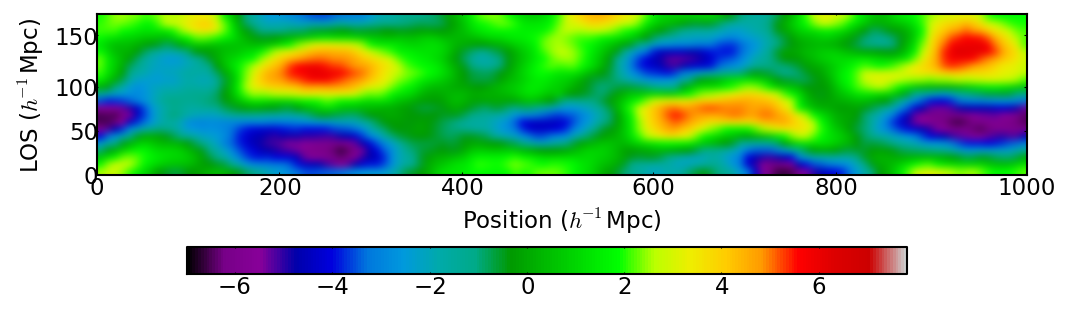}
  \caption{Impact of foreground cleaning on the matched-filtered field. This is similar to Figure \ref{fig:WienerLOS}, except
that the results here are for a matched filter with a template 
radius of $R_{\text{T}} = 35 \hmpc$.}
  \label{fig:MatchedLOS}
\end{figure}

With the estimates of the previous section as a rough guide, we now
apply the matched filter to our noisy mock redshifted 21 cm data.
In order to illustrate the results of passing our mock data through
a matched filter, we start by examining simulated signal-to-noise fields
for a single template radius of $R_{\text{T}} = 35$ $h^{-1}$Mpc. This template
radius corresponds to the typical size of the ionized bubbles we believe
we can detect (see Figure \ref{fig:ToySNR}). A representative slice
through the simulation is shown in Figure \ref{fig:ThreeFigureMatched}.
The results look promising, with signal-to-noise ratios comparable
to the values anticipated in the idealized calculation of Figure \ref{fig:ToySNR}. 
Although the Wiener filter provides
the best overall map, or data cube, one can still detect individual features at greater
significance by applying a matched filter.
Comparing
with Figure \ref{fig:Wiener}, it is clear that the Wiener filter is
passing more small scale structure than the matched filter shown here.
This results in the signal-to-noise ratio being larger (in absolute value) 
for the matched filter than for the Wiener filter.
In particular, we find values
of the signal-to-noise ratio that are as low as $\sim -10$ in the matched filter data cube, a significant improvement over the global minimum 
of $\sim -6$ for the Wiener filter.
Moreover, we can compare the filtered noisy signal in the bottom-left panel
with the filtered pure noise field in the top-right panel. They differ by
more than in the case of the Wiener filter. Indeed, the very low signal to noise
ratio regions (shown in dark blue/purple in the bottom-left panel) line up fairly well
with ionized regions in the top-left panel. This is especially apparent when comparing the filtered noisy signal to the filtered noiseless signal, shown in the bottom-right panel. For the slice shown, almost all of the significant features in the filtered noiseless signal are preserved in the noisy case. Figure \ref{fig:MatchedLOS}
shows the impact of foreground cleaning: as in the case of the Wiener filter (Figure \ref{fig:WienerLOS}),
this compresses structures along the line of sight and reduces the overall
signal-to-noise ratio in the data cube. The signal-to-noise ratio is still 
significant enough,
however, to robustly identify ionized regions.

As with the Wiener filter,
and in what follows subsequently, we show slices of $8$ $h^{-1}$Mpc thickness. This
choice is arbitrary, but we expect similar results provided the slice thickness
is small compared to the radius of the template filter. It is important to keep
in mind, however, the full data cube will consist of many separate slices
of this thickness. Also note that the transverse dimension of the MWA-500 is larger
than that of our simulation box by a factor of $\sim 3$, and so these slices represent only $\sim 1/9$
of the MWA field of view.

These results are promising, but they are for a single filtering scale, and
so we can 
do significantly better by considering a range of template radii, and looking
for extrema in the resulting signal-to-noise fields. In particular,
we proceed to apply a sequence of filters with template radii up to
$R_{\text{T}} \leq 75\hmpc$ -- see \S \ref{sec:VaryRt} for a justification of
this maximum -- across the simulation volume. 
We assign the minimum (most negative)
signal-to-noise value obtained over the range of template
radii to each simulation pixel and use this to construct a new field. The position of local minima
in this field are chosen to be the centers of candidate bubbles, and each such bubble is
assigned a radius according to the scale of the template filter that minimizes its
signal-to-noise. We focus on \textit{minimum} values since ionized regions are expected
to appear as regions of low 21 cm signal. All candidate bubbles whose central signal to noise
is lower than $-5$ are considered to be detected ionized regions.

We find it important to apply one additional criterion to robustly
identify ionized regions. The criterion is that a low signal to noise
region on scale $R_{\text{B}}$ must additionally be {\em low in signal to noise at all smaller smoothing scales},
$R_{\text{T}} < R_{\text{B}}$. This guards against the possibility that a detected
bubble will be centered on neutral material that is nevertheless surrounded by ionized hydrogen. A
region like this will have a high (least negative) signal-to-noise when filtered on small scales
and then dive down (gaining statistical significance) when filtered on scales containing
the surrounding ionized material. We discard such spurious bubbles by requiring
that the field is low on all smaller smoothing scales. The only downside to this
procedure is that it occasionally discards true ionized regions whose center
happens to coincide with a significant upward noise fluctuation. Overall, however, it improves the quality of detected
bubbles (\S \ref{sec:success}). This cut also requires a threshold choice; we
reject candidate regions if their signal-to-noise ratio crosses above a threshold $\mathcal{S}_{\max}$
at any smoothing scale less than $R_{\text{T}}$.
After trying several
thresholds, we found the most effective choice to be $\mathcal{S}_{\max} = -1\sigma$. In principle, one might use the full curve of signal-to-noise ratio
versus template radius for each candidate bubble to help verify the detection and
determine the properties of the bubble. In practice, we found that individual
signal-to-noise curves are noisy and difficult to incorporate into our analysis and
so we don't consider this possibility further in what follows.

\begin{figure*}
  \centering
  \includegraphics[width=16.8cm]{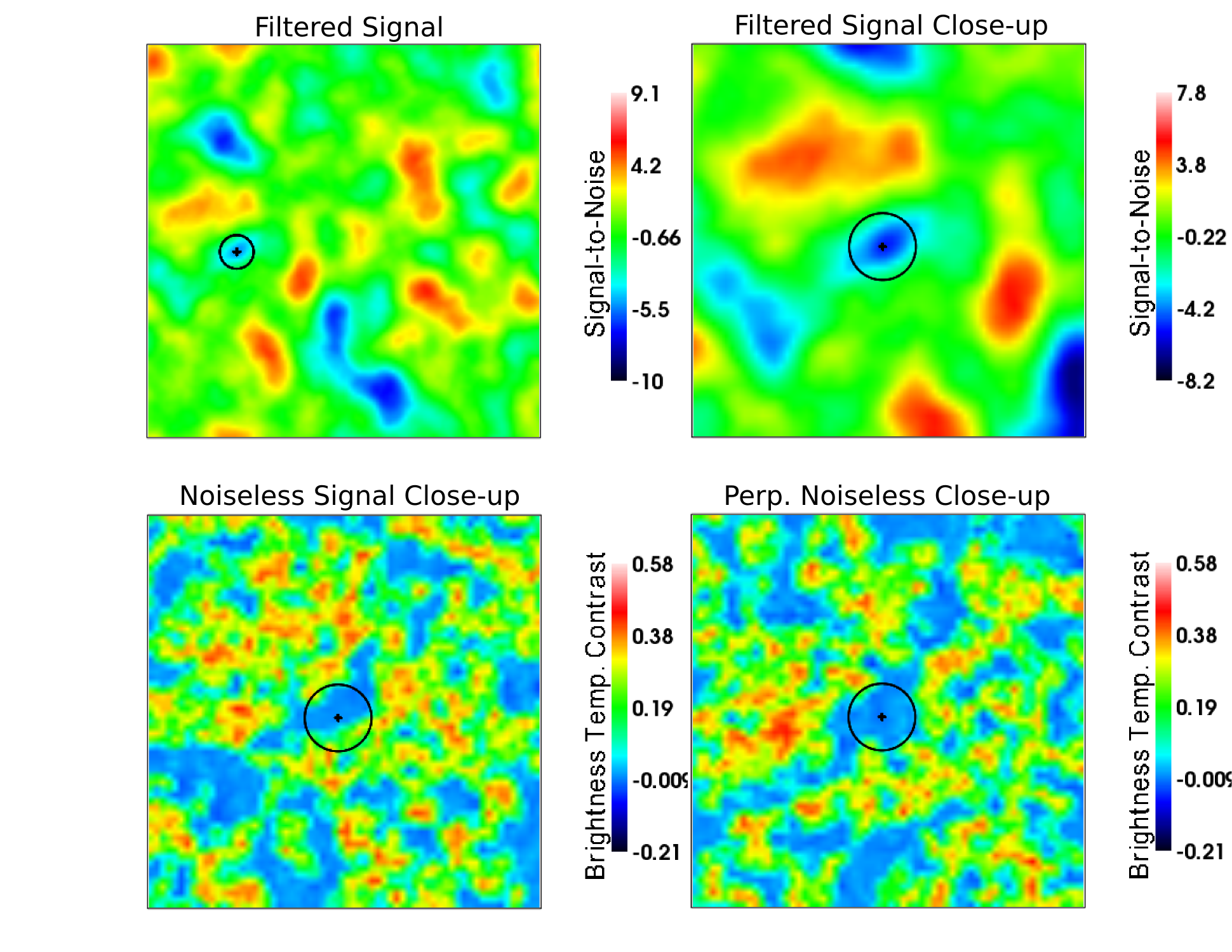}  
  \caption{An example of a detected ionized region. \textit{Top-left:}
    Signal-to-noise field after applying the matched filter to
    the noisy signal. The detected bubble is plotted on top of the
    corresponding region in the map. \textit{Top-Right:} Zoomed-in
    view of the detected bubble in the matched-filtered
    map. \textit{Bottom-Left:} Detected bubble superimposed on a
    zoomed-in view of the noise-less unfiltered 21 cm brightness
    temperature contrast map. \textit{Bottom-Right:} A perpendicular
    zoomed-in view of the bubble depicted in the bottom-left
    panel. All matched-filtered maps use the template radius that
    minimizes the signal-to-noise at the center of the detected
    bubble. In the top-left case, the boxlength is $L = 1\hgpc$,
    while in the zoomed-in slices it is $L \approx 500\hmpc$.}
  \label{fig:ShowAverageBubble}
\end{figure*}

We apply this algorithm to the mock redshifted 21 cm data and
identify $220$ ionized regions across the simulation volume (which is different
than the MWA survey volume, as we will discuss subsequently). A representative
example of a detected bubble is shown in Figure \ref{fig:ShowAverageBubble}. The
circle in the figure identifies the detected bubble size and the location
of its center in both the filtered noisy signal (top-left and top-right panels),
as well as in the input signal (bottom-left and bottom-right panels). The
algorithm has convincingly identified an ionized region. The detected bubble overlaps
a small fraction (\%10) of neutral material in the input signal. Although this particular
ionized bubble is well identified, most of the ionized regions in the signal will
escape detection. This is because the significance levels of the detected bubbles are not that 
high, and an ionized region generally needs to be coincident
with a downward fluctuation in the noise to pass our significance threshold. 
For example, consider the larger ionized region below and to the left of the
detected region in the bottom-left panel of Figure \ref{fig:ShowAverageBubble}.
This region, while larger and therefore more detectable on average than the identified bubble,
happens to coincide with a large upward noise fluctuation and hence fails to cross
the significance threshold. While we can not identify all of the large
ionized regions in the noisy mock data, we can robustly identify some regions; this may still
be quite valuable.

It is also clear that the underlying ionized regions are manifestly non-spherical,
creating an ambiguity as to what the appropriate `radius' of the
region is. Focusing on the bottom right panel in
Figure \ref{fig:ShowAverageBubble}, we could imagine the size being
reasonably described by a radius $\sim \%50$ larger, so as to enclose more
of the nearby ionized material. However, our method naturally favors
radii causing little overlap with neutral material at these size
scales. Therefore, an ionized region like the one shown in
Figure \ref{fig:ShowAverageBubble} is more likely to be detected as several
small ionized regions than one large one, although both
characterizations seem reasonable. 

\begin{figure}[h]
  \centering
  \includegraphics[width=8.4cm]{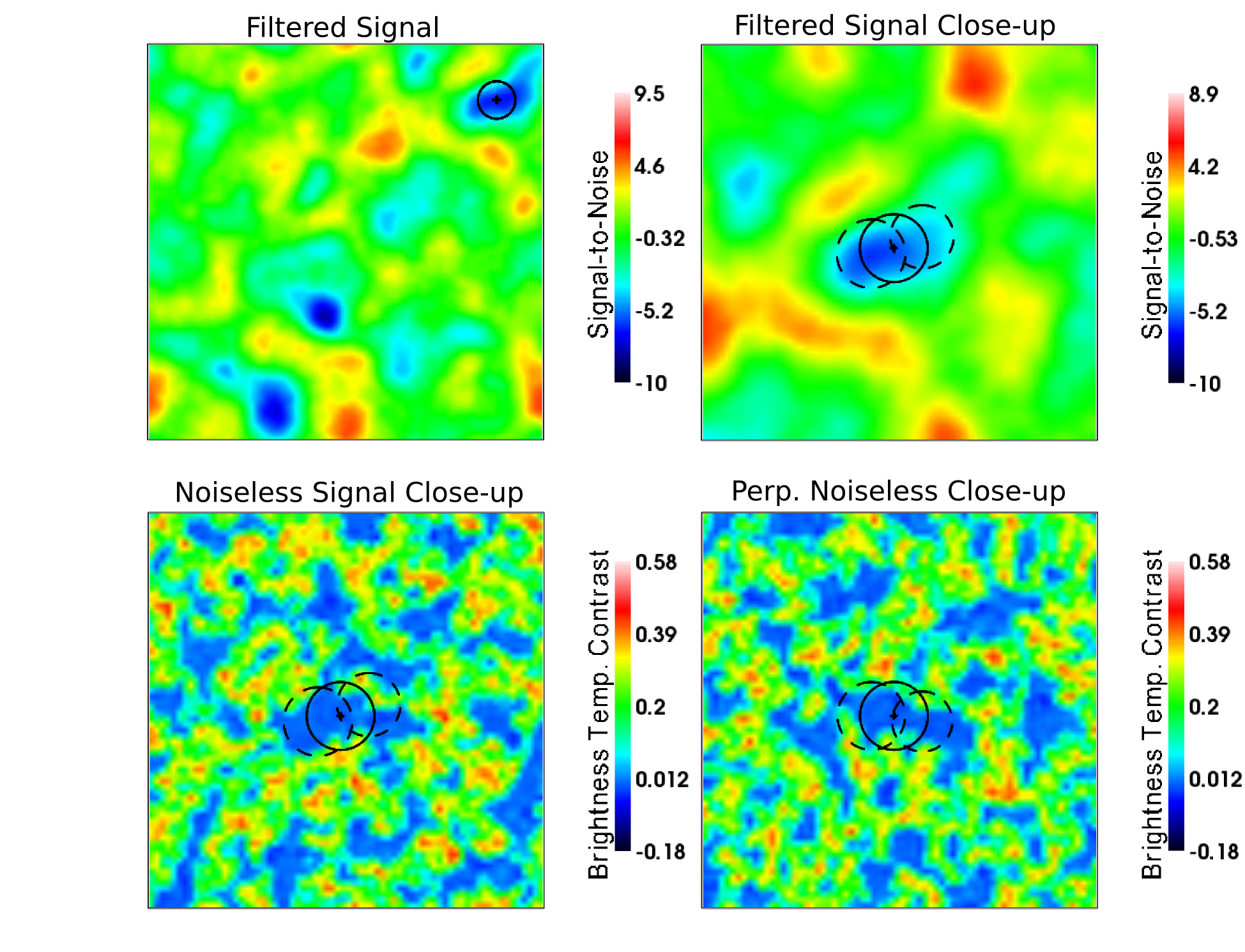}  
  \caption{An example of an ionized region that our algorithm detects as
several neighboring bubbles. \textit{Top-left:} Signal-to-noise field
    after applying the matched filter to the noisy signal. The
    main detected bubble is plotted on top of the corresponding region
    in the map. \textit{Top-Right:} Zoomed-in view of the main
    detected bubble in the matched filtered map (solid curve) along
    with two other nearby detected bubbles (dashed
    curve). \textit{Bottom-Left:} The detected bubble superimposed on the
    zoomed-in, noise-less, unfiltered 21 cm brightness temperature
    contrast map. Again, the additional nearby detected bubbles are
    shown (dashed curve). \textit{Bottom-Right:} A perpendicular view
    of the bubble depicted in the bottom-left panel, with the nearby
    detected bubbles visible. All matched-filtered maps use the
    template radius that maximizes the signal to noise at the center
    of the main detected bubble. The box length in the top-left figure
    is $L = 1\hgpc$, while in the zoomed-in panels, the box length
    is $L = 550 \hmpc$.}
  \label{fig:ShowBubbles}
\end{figure}

Figure \ref{fig:ShowBubbles} gives a further example of how the algorithm identifies
bubbles, and some of the ambiguities that can result. This figure
shows an example of an irregular, yet contiguous, ionized region that is
detected as more than one ionized bubble.
Here we show spatial slices through the
center of the middle sphere, marked with a solid circle, which happens to intersect
neighboring ionized bubbles, whose cross sections are shown as dashed circles.
Hence, our algorithm generally represents large, irregularly shaped, yet contiguous,
regions as multiple ionized bubbles.

It is important to emphasize further the difference between the
simulated results shown here and the idealized test case of the
isolated bubble shown in the previous section. In particular, we
consider here the application of matched filters to the 21 cm signal
during a late phase of reionization  in which many ionized regions,
with a broad size distribution, fill the survey volume: the ionized
regions {\em are not} isolated bubbles in a sea of partly neutral
material. When applying a matched filter of template radius
$R_{\text{T}}$ around a point, the values of the field at many
neighboring pixels impact the filtered field at the point in
question. It is hence possible that a filtered pixel is affected by
several distinct neighboring ionized regions. Indeed, this can result
in even neutral regions having low signal-to-noise ratios provided
they are surrounded by many nearby ionized regions. For instance, in
the low noise limit, {\em any region with volume-averaged neutral
  fraction lower than the cosmic mean would pass our significance
  threshold}.  To guard against this type of false detection, we
implemented the requirement that a candidate bubble has low signal-to-noise 
for {\em all} template radii smaller than the detected
radius. Another possibility might be to treat small ionized regions as
an additional noise term in the filter. However, in practice, our attempts along
these lines introduced an additional level of model dependence without
significantly increasing the quality of the
detected bubbles. Ultimately, it is important to keep
in mind that the signal-to-noise values quoted here reflect only the
likelihood that a value arises purely from noise, and so they
are not strictly indicative of the quality of the
detected bubbles.

\subsection{Success of Detecting Ionized Regions} \label{sec:success}

We hence turn to describe the characteristics of the detected ionized regions,
and to quantify the method's level of success in detecting ionized bubbles. To do this, we
calculate the fractional overlap of each detected bubble with
ionized material in the underlying signal. Additionally, we estimate how
many ionized bubbles should be detectable across the entire MWA-500
survey volume.

The matched filter technique finds $220$ bubbles across our simulation
volume. However, the algorithm for determining bubble positions and sizes allows for bubbles to occupy overlapping areas, as shown in Figure \ref{fig:ShowBubbles}. We find that $\sim 55\%$ of the detected bubbles have \textit{some} overlap with another bubble, although only $\sim 15\%$ of the {\em total} volume occupied by detected bubbles is occupied by more than one.  
Regardless, $\sim 96\%$ of the detected ionized bubbles have an average ionized
fraction larger than $x_i = 0.79$, which is the volume-averaged ionization
fraction of the simulation box at the redshift under
consideration. Furthermore, $\sim 42\%$ of the detected bubbles have an
ionized fraction greater than $x_i = 0.9$. 
The lowest ionization fraction of a detected bubble is
$x_{i,\min} = 0.77$, just slightly
below the volume-averaged ionization fraction of the simulation. In
total, we detect 9 bubbles whose ionized fractions are less than the
average ionization fraction of the box. Inspection reveals that
these regions happen to be coincident with significant ($\leq -3-\sigma$) 
downward noise fluctuations.

\begin{figure}[t]
  \centering 
   \includegraphics[width=8.8cm]{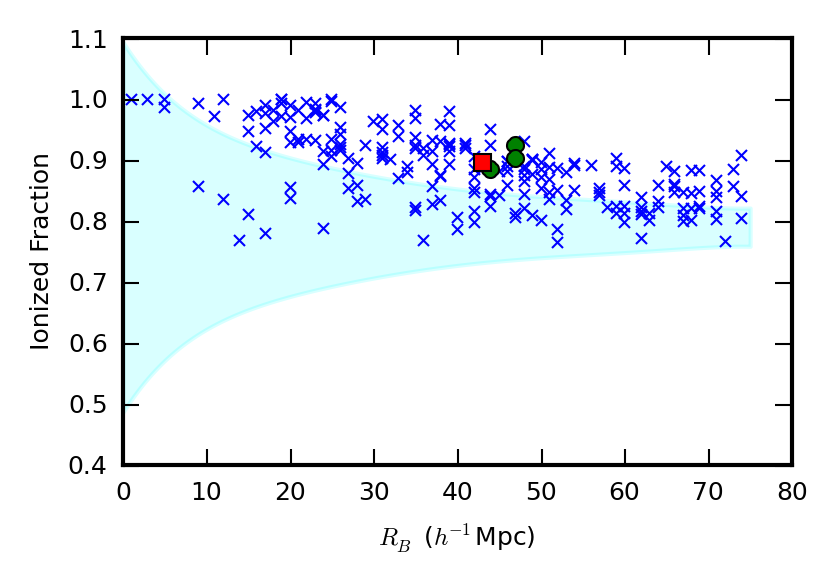}
  \caption{A measure of the bubble detection success rate. The points ($\times$) show 
    the volume-averaged ionized fraction of detected bubbles versus
    their detected radius. For comparison, the cyan shaded region shows the 1-$\sigma$ spread
in the ionized fraction of {\em randomly} placed bubbles of the same radii.
 The bubble 
depicted in \Fig{fig:ShowAverageBubble} is
    marked with a large red square, while the three bubbles shown in
    \Fig{fig:ShowBubbles} are marked with large green circles.}
  \label{fig:success}
\end{figure}

In Figure \ref{fig:success}, we plot the volume-averaged ionized fraction within each of our
detected bubbles against
the detected bubble radius for a significance threshold of $5\sigma$. For comparison, we show
the $1-\sigma$ spread in the ionized fraction enclosed by {\em randomly} distributed
spheres of the same size.\footnote{The $1-\sigma$ spread shown in the figure extends past $x_i = 1$, but this is only because the distribution of ionized fractions is not symmetric about the mean, i.e., the probability
distribution function of the ionized fraction is non-Gaussian.} The spread in ionization of the randomly distributed spheres around
the box average ionized fraction, $\avg{x_i} = 0.79$, decreases with increasing radius; this
reflects the drop off in the power spectrum of the ionization field towards large scales.
Most of the detected bubbles 
are significantly more ionized than random regions, as expected, indicating
a significant success level. There are a few poor detections which result mostly from
downward noise fluctuations. There is a small overall decrease in the ionized
fraction of detected bubbles larger than $R_{\text{B}} \gtrsim 40$ $h^{-1}$Mpc, suggesting that
we may no longer be detecting individual ionized regions here. These
regions may potentially be distinguished from isolated bubbles by examining
the signal-to-noise ratio as a function of template radius closely, as
we discuss in \S\ref{sec:VaryRt}.

We can estimate the number of ionized regions detectable in the MWA-500
by scaling from our simulation volume to the MWA survey volume.
At $z_{\rm fid} = 6.9$, for an ionized fraction of $\avg{x_i} = 0.79$,
we expect to find $140$ bubbles in a $B=6 \MHz$ chunk of the
MWA, over its entire field of view of $\sim 770$ deg$^2$.
About $135$ ($60$) of these detected bubbles are expected to
have ionized fractions larger than $79\%$ ($90\%$). 
This estimate
comes from simply scaling our simulation volume (which is deeper
than the MWA bandwidth) to a $6 \MHz$ portion of the MWA survey volume. 
Analyzing the MWA data over a $6 \MHz$ 
chunk is meant to guard against redshift evolution: the full bandwidth
of the survey is $B = 32 \MHz$ and so the prospects for bubble detection
across the full survey are even better than this estimate suggests.
The precise gain will be dependent on how rapidly the bubble size distribution
evolves across the full survey bandwidth. One caveat with our estimate, however,
is that $B=6 \MHz$ corresponds to only $\sim 70\hmpc$. This is comparable
to the size of our larger bubbles, and so analyzing chunks this small
might weaken our ability to detect large bubbles. This effect
is not incorporated in our scaling estimate, which simply
takes the ratio of the MWA survey volume and our simulation volume.
In practice, one can perform the bubble extraction for different
analysis bandwidths to help ensure robust detections.

By increasing the significance threshold for detection, one might hope to improve the quality of the detected bubbles. However, we find that this leads to the detected bubbles being {\em larger} on average rather than being more ionized. This results from the fact that, at a fixed ionized fraction, a larger bubble will have a greater (in magnitude) signal-to-noise value than a smaller bubble. By increasing the significance threshold, some smaller, highly-ionized bubbles are no longer detected, while larger, less-ionized bubbles are. Specifically, we find that by increasing the threshold to $6-\sigma$ ($7-\sigma$) the number of detected bubbles drops to $\sim 100$ ($\sim 50$) with 99\% (100\%) being more ionized than the box, but only 34\% (33\%) being more than 90\% ionized. Hence, for
any of these thresholds we robustly detect many bubbles, although the detailed
success rate and overall number varies somewhat with the precise choice of threshold. Note that lowering the significance threshold has the undesirable effect of increasing the number of false positives. Ultimately, with real data one should
explore a range of threshold choices and examine the impact.

\subsection{Range of Template Radius Considered} \label{sec:VaryRt}

It is worth mentioning one further detail of our algorithm. In
the previous section, we set the maximum template radius considered at
$R_{\rm T, max} = 75$ $h^{-1}$Mpc, without justification. In fact, we
have a sensible and automated way for arriving at this choice. We discuss this procedure briefly here.
 
A good candidate ionized region should in fact obey three criteria. First,
it should have a large (negative) signal-to-noise ratio, so that it is unlikely
to result from a noise fluctuation. Second, the signal-to-noise ratio should
be small for all trial radii smaller than the optimal template radius, as discussed
in \S \ref{sec:mapplied}. Finally,
the total signal must itself be
small in an absolute sense. In the limit of low noise, anything less neutral 
than average would qualify as a bubble by the first criterion, and so
this third criterion may then become important for robustly identifying bubbles.
This low noise limit is relevant for the MWA-500 only on very large
smoothing scales, where the noise averages down significantly. 

Since this third criterion becomes important only on very large smoothing
scales here, we use it only to set the maximum template radius considered.
Without this third consideration, our algorithm generally identifies a few
excessively large ionized bubbles, but this can be easily understood
and avoided as follows. Consider, for the moment, the 21 cm brightness temperature
field in the absence of noise and foregrounds. Let's
further work in units of $T_0$ (Equation \ref{eq:21cm}), and
remove the average brightness temperature contrast across the data cube.
In this case, the signal inside a highly ionized bubble is 
expected to be $-\avg{x_{\rm HI} (1 + \delta_\rho)}$. If we now spherically average the field on scales 
smaller than the bubble, the value at bubble center will not change from 
this value, $-\avg{x_{\rm HI} (1+\delta_\rho)}$. Once
the smoothing scale becomes larger than the bubble scale, however, 
surrounding neutral
material will increase the value of the filtered field at bubble center. 
Hence, if the filtered field becomes everywhere larger than
$-\avg{x_{\rm HI} (1 + \delta_\rho)}$ on some smoothing scale, it is clear
that no larger ionized bubbles exist within the data cube. 
This suggests that we can set the maximum template radius 
by requiring that the filtered noisy signal reaches sufficiently
small values, at some locations across the data cube, for there to still plausibly be
completely ionized regions. Since the presence of noise only increases
the variance, this should provide a conservative estimate of the
maximum size of the ionized regions. In practice, we need to chose a threshold criterion without
assuming prior knowledge of the neutral fraction.
Here we set the maximum template radius to be the smallest smoothing
scale at which the filtered noisy field everywhere exceeds $-\avg{x_{\rm HI} (1 + \delta_\rho)} \geq - 0.075$. This corresponds to the expected contrast at $\axhi = 0.1$, assuming $\avg{\delta_x \delta_\rho}=-0.25$, and yields a maximum template radius of 
$R_{\text{T,max}} = 70\hmpc$, $73\hmpc$, $75\hmpc$, and
$75\hmpc$ for $\axi = 0.51$, $0.68$, $0.79$, and $0.89$, respectively.
The precise threshold used here, $-0.075$, is somewhat arbitrary but this choice 
is only being
used to set the maximum template radius considered.\footnote{This choice
might appear to preclude the possibility of detecting bubbles at the end of reionization when $\avg{x_{\rm HI} (1 + \delta)} \leq 0.075$. However, the
threshold choice is only used to set the maximum template radius, and
so ionized regions may still in principle be detected at these late
stages of reionization. The ionized regions identified 
at the end of reionization
are, however, generally less robust 
given the reduced contrast between fully ionized  
and average regions at these times (see \S \ref{sec:VaryXi}).}

\section{Variations on the Fiducial Model} \label{sec:Variations}

So far, we have considered the prospects for bubble detection only
in our fiducial model with $\avg{x_i} = 0.79$ at $z_{\rm fid} = 6.9$ and
only for the MWA-500. Here we consider first alternate models in
which the Universe is more or less ionized at $z_{\rm fid} = 6.9$ than in our
fiducial case, and then consider how the sensitivity declines towards
higher redshifts at fixed ionized fraction. In addition, we consider variations
around our fiducial assumptions regarding the impact of foreground cleaning.
Then we turn to consider
the sensitivity of the MWA-128; this is meant to illustrate the
prospects for bubble detection with the very first generation of redshifted
21 cm surveys, while the MWA-500 represents a second generation survey.

\subsection{Ionized Fraction} \label{sec:VaryXi}

In order to consider bubble finding at earlier and later stages of the
EoR, we run our matched filter on simulation outputs with volume-averaged 
ionized fractions of $\avg{x_i} = 0.51, 0.68$, and $0.89$. As discussed in
\S\ref{sec:wiener} these outputs are actually at slightly different redshifts,
but we generate the 21 cm field as though they were at $z_{\rm fid} = 6.9$.
As far as bubble detection is concerned, varying the ionized fraction leads
to two, generally competing, effects. First, the bubbles grow as reionization
proceeds. This tends to boost detection, since it is only the large
scale modes that are detectable over the thermal noise. Second, however,
the {\em contrast} between an ionized region and a typical volume
of the Universe is reduced as reionization proceeds. This makes bubble
detection more difficult. Both of these effects are quantified in
the idealized isolated bubble case in Figure \ref{fig:ToySNR}.
It is also clear that the ideal ionized fraction for bubble detection
will be somewhat survey dependent. As already illustrated in
Figure \ref{fig:ToySNR} and discussed further in \S \ref{sec:LOFAR},
a LOFAR-type interferometer will perform better when the ionized
regions are smaller. 

\begin{figure}[h]
  \centering 
  \includegraphics[width=4.2cm]{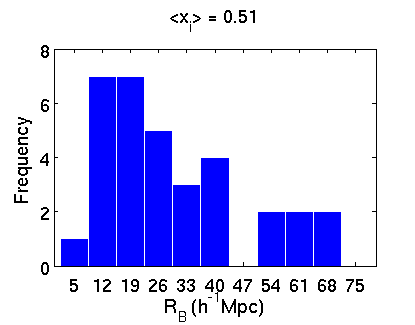}
  \includegraphics[width=4.2cm]{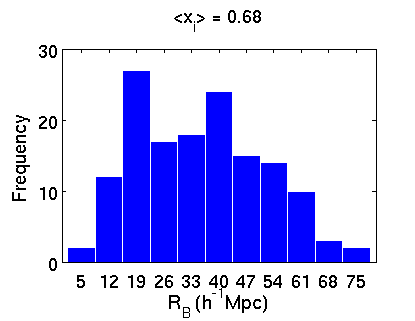}
  \includegraphics[width=4.2cm]{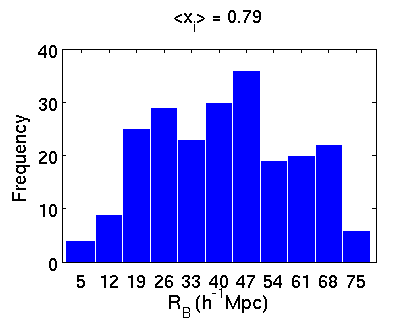}
  \includegraphics[width=4.2cm]{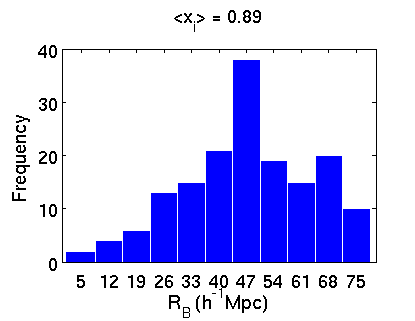}
  \caption{Size distributions of detected bubbles for varying (volume-averaged) ionization fractions. 
    The histograms show the size distribution of (identified)  ionized regions 
for simulation snapshots with
    volume-averaged ionized fractions of $\avg{x_i} = 0.51$ (top-left),
    $0.68$ (top-right), $0.79$ (bottom-left), and $0.89$
    (bottom-right). These figures demonstrate how the total number and size
    distribution of detected bubbles varies with ionized fraction.}
  \label{fig:SizeHists}
\end{figure}

We find that the matched filter is capable of detecting ionized
regions for each of the ionized fractions studied. In
Figure \ref{fig:SizeHists} we show histograms of the detected
bubble size distributions for each ionized fraction. Since
we preferentially detect large ionized regions, we don't
expect these distributions to be representative of the
true underlying bubble size distributions. For example, in Figure
\ref{fig:SizeHists}, the size distribution peaks around $\gtrsim40\hmpc$ for
the case of $x_i = 0.79$, despite volume-weighted size distribution peaking around $\sim 30\hmpc$ in
\href{http://iopscience.iop.org/0004-637X/654/1/12/fulltext/65071.fg4.html}{Figure
  4} of \cite{Zahn:2006sg} at roughly the same ionized fraction. Nonetheless,
the histograms illustrate a general shift from smaller to larger detected
bubble radius as the ionized fraction increases. By applying
the matched filter to several redshift bins, one can potentially
observe precisely this evolution with the MWA-500. This would
complement studies of the 21 cm power spectrum evolution over
the same redshift range (e.g., \citealt{Lidz:2007az}). From
the histograms, one can see that -- of the models shown -- the 
best ionized fraction
for bubble detection is $\avg{x_i} = 0.79$. This is apparently near the sweet spot for the MWA-500 where
the bubbles are large enough in the model for detection, but
the contrast with typical regions is still sufficiently large.

The average ionized fraction within detected bubbles varies
significantly across the different ionized fractions considered.
Specifically, the percentage of detected bubbles that are more than
90\% ionized is $0\%$, $15\%$, $43\%$, and $91\%$ for $\axi = 0.51$,
$0.68$, $0.79$, and $0.89$, respectively.  However, in each case the
percentage of detected bubbles with ionized fraction larger than the
(global) volume-averaged ionization fraction is fixed at $\gtrsim 95\%$. At
first glance, one aspect of these results may appear to be in tension
with the calculations of \S\ref{sec:toy}, where we estimated that
bubble detection would be unsuccessful for neutral fractions larger
than $\avg{x_{\rm HI}} \gtrsim 0.5$.  However, this estimate
considered the detection of {\em isolated} bubbles. Inspection reveals
that the detected bubbles at $\axi = 0.51$ each correspond to clusters
of smaller ionized regions. Evidently, these appear as a single larger ionized
region after convolving with the template filter and downweighting
the noisy short-wavelength modes. In practice, it may be possible to distinguish
this case from that of an isolated bubble by analyzing the trend of 
signal-to-noise ratio versus trial template radius. The signal-to-noise
ratio is expected to grow more rapidly with radius (before reaching the bubble
scale) for a truly isolated
bubble.

\subsection{Timing of Reionization} \label{sec:Timing}

We now consider how the prospects for bubble detection diminish if reionization occurs
earlier and the observations are focused on the corresponding redshifts.
In particular, we examine the case that our model with an ionized fraction
of $\avg{x_i} = 0.79$ is observed at a higher redshift. We focus on this case since
this ionized fraction appears close to optimal for bubble detection.
Aiming for only a rough estimate here, we consider the prospects for
detecting a $R_{\text{B}} = 40$ $h^{-1}$Mpc bubble. 

Although several different factors in the noise power spectrum of
Equation \ref{eq:NoisePower} scale with redshift, the dominant
scaling is with the sky temperature. The noise power scales
as $P_{\text{N}} \propto T_{\text{sky}}^{2}$, and the
sky temperature follows $T_{\text{sky}} \propto \nu^{-2.6} \propto
(1+z)^{2.6}$. Therefore, we expect the signal-to-noise ratio of
a detected bubble to fall off with increasing observation redshift roughly as
\begin{equation}
\mathcal{S}(z) = \mathcal{S}(z_{\text{fid}}) \left(\frac{1+z_{\text{fid}}}{1+z}\right)^{2.6}. \label{eq:ZDependence}
\end{equation}
This indicates the signal-to-noise ratio for a bubble detected with
a signal-to-noise of $\mathcal{S}(z_{\text{fid}})$ at redshift $z_{\text{fid}} = 6.9$, 
if the bubble were instead observed at redshift $z$. A relatively large bubble with
$R_{\text{B}} \approx 40$ $h^{-1}$Mpc has a typical signal-to-noise ratio at bubble center 
of $\mathcal{S} \approx 4$
at our fiducial redshift. This value is found by incorporating foreground cleaning into the corresponding curve in Figure \ref{fig:ToySNR}. 
According to Equation \ref{eq:ZDependence}, the signal-to-noise value will be reduced to a significance of $\mathcal{S}
\approx 2$ (1) at $z = 9.3$ (12.5). The bubble will, in fact, be more detectable than implied by this one number -- the
signal-to-noise ratio at bubble center -- since
an ionized region should have low signal-to-noise over much of its volume. From this, we conclude that bubble
detection should be feasible with the MWA-500 if our fiducial ionized fraction occurs later than $z \lesssim 9$ or so,
but that the prospects are rather limited in the case of significantly earlier
reionization. A range of recent work in the literature, however, suggests that
reionization is unlikely to complete so early. See, for example, Figure 9 from the recent
study of \cite{Kuhlen:2012vy} which combines
Ly-$\alpha$ forest data (\citealt{Fan:2005es}), measurements of the
Thomson optical depth from WMAP (\citealt{Komatsu:2010fb}), and measurements of the Lyman-break
galaxy luminosity function (\citealt{Bouwens:2009qs}). Hence, the prospects for bubble detection appear good for the MWA-500.

\subsection{Effects of Foreground Cleaning} \label{sec:ForegroundCleaning}

Next, we consider the impact of variations around our standard foreground
cleaning model. As discussed previously (\S \ref{sec:foregrounds}), our
standard assumption is that the impact of foreground cleaning
can be approximately mimicked by removing the running mean, over a bandwidth
of $B=16$ MHz, across each line of sight. The optimal foreground cleaning
strategy avoids `over-fitting' by removing the smoothest possible function
over the largest possible bandwidth, in order to preserve the underlying signal
as much as possible. It also avoids `under-fitting' by ensuring that foreground
residuals do not excessively contaminate the signal. \cite{Liu:2011ih}, for
example, find that 21 cm foregrounds can be removed to one part in
$10^5$ or $10^6$ by subtracting roughly three modes over $\sim 32$ MHz of
bandwidth. This should have a fairly similar impact to our fiducial
cleaning model, but we would expect a bit more degradation in this case.
A detailed investigation would add foreground contamination into our
mock data cubes, and explore the impact of various cleaning algorithms directly.


Here, we instead check how our results change for slightly more
and less aggressive foreground cleaning. In particular, we remove
the running mean over each of $B=32$ MHz, $8$ MHz, and $6$ MHz and rerun our
bubble finding algorithm (at $z_{\rm fid} = 6.9$, $\avg{x_i} = 0.79$). This has
little impact on our results for the case of $B = 32$ MHz and $8$ MHz. In particular, the number of identified
bubbles varies by less than $10-15\%$ and the quality of detected bubbles decreases slightly
for the more aggressive cleaning model, and improves slightly in the most optimistic
case. Specifically, for $B=16$ MHz, $96\%$ ($43\%$) of bubbles have
ionized fractions exceeding $x_i = 0.79$ ($0.9$); for $B=8$ MHz the
corresponding numbers are $90\%$ ($32\%$); and for $B=32$ MHz
the same numbers are $96\%$ ($62\%$). For the most aggressive case of $B=6$ MHz, we find the number of detected bubbles drops significantly to $\sim 50\%$ of the number detected in the fiducial case. The larger degradation in the $B=6$ MHz results because the foreground cleaning
scale has become comparable to the scale of many of the detected bubbles.
Despite this drop, we find the quality of the detected bubbles to be about the same as with the $B = 8$ MHz model, with $92\%$ ($36\%$) exceeding $x_i = 0.79$ ($0.9$). 
The distribution of detected bubble sizes also depends on the bandwidth used for foreground cleaning, with smaller  bandwidths generally corresponding to smaller detected bubble sizes when $B \lesssim 16$ MHz. This behavior is consistent with work by \cite{Petrovic:2010me} who found that a bubble whose size is comparable to the scale for foreground cleaning should have its contrast reduced. Even in the most aggressive case considered, however, we still detect many
ionized regions robustly. While these estimates are encouraging,
a more detailed study is warranted. It may also be advantageous to
estimate the power spectrum of the foregrounds, and incorporate this
as an additional noise term in Equation \ref{eq:WienerFilter} and Equation \ref{eq:MatchedFilter} for each of the Wiener filter and the matched filter, respectively.


\subsection{128 Antenna Tile Configurations} \label{sec:128tiles}

So far our analysis has focused on the MWA-500, which is meant
to represent a second generation 21 cm survey. In the near term, it
is timely to consider the prospects for a 128 tile version
of the MWA (the MWA-128) which is ramping up to take data in 
the very near future. This should be significantly less
sensitive, since the number of baselines scales as the
number of antenna tiles squared.

In order to generate thermal noise representative of
the MWA-128, we start by considering a similar antenna
distribution as for the MWA-500. In particular, we
assume all of the antenna tiles are packed as closely
as possible within a core of radius $8 \meter$ and that
the antenna distribution subsequently falls off as
$r^{-2}$ out to a maximum baseline of $1.5$ km. After
comparing the thermal noise power spectrum in this
configuration with that in \cite{Beardsley:2012za}, we
find that our noise power is larger by up to a
factor of a few.
This could possibly be due to our approximation of a
smooth antenna distribution being less valid for the MWA-128, or to the 
fact that our analytic formula
for the noise power spectrum
does not incorporate a full treatment of rotation synthesis. In
an effort to bracket somewhat the impact of the detailed antenna
distribution, we further consider a configuration where all antenna
tiles are packed as closely in a dense core of radius $\sim 25 \meter$.
This resembles the `super-core' configuration considered in \cite{Lidz:2007az}
for the power spectrum.

\begin{figure}[h]
  \centering
  \includegraphics[width=8.4cm]{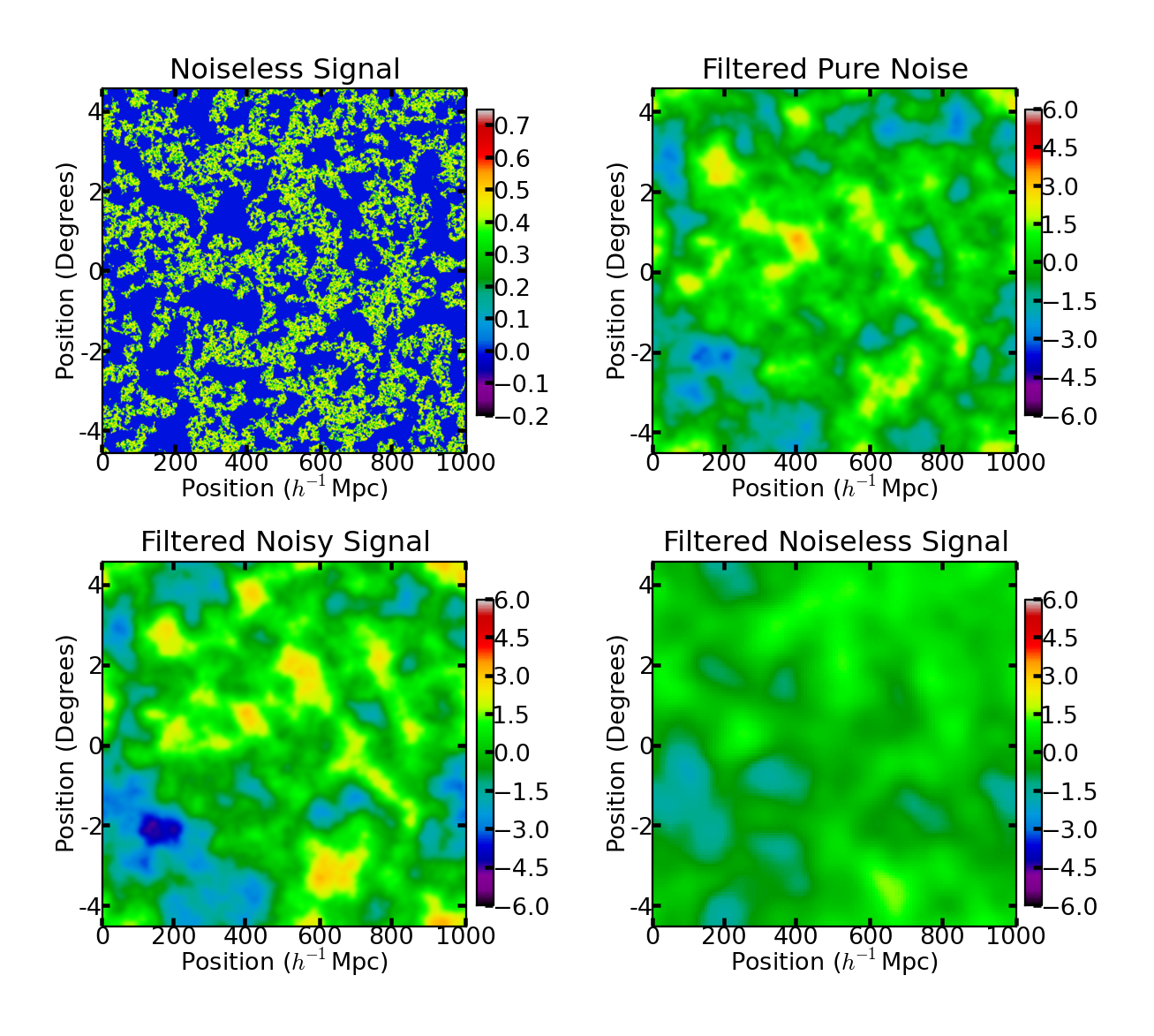}
  \caption{Bubble detection with the MWA-128. This figure is similar to Figure \ref{fig:ThreeFigureMatched}, except it is for the MWA-128 configuration rather
than for the MWA-500.}
  \label{fig:128tiles}
\end{figure}

The results of applying the optimal matched filter for a
single template radius of 35$h^{-1}$Mpc are shown in Figure
\ref{fig:128tiles} for the $r^{-2}$ tile distribution. This
shows that the sensitivity is much lower than for the MWA-500, as
expected. It is much more difficult to distinguish the filtered noisy
signal (bottom-left panel) from the filtered pure noise (top-left) panel here than in 
Figure \ref{fig:ThreeFigureMatched}. Most of the significant, dark blue regions in the filtered noisy
signal correspond simply to low noise regions. However, applying the detection algorithm
we do nonetheless detect $7$ bubbles across a volume equivalent to a $6 \MHz$ chunk of the MWA-128 survey.
The success is generally lower than in the case of the MWA-500: here
75\% of detected bubbles exceed the average
ionized fraction of the box, while $\sim 42\%$ exceed $x_i =
0.9$. In the supercore configuration, we find slightly higher significance levels (up to $6.9-\sigma$)
but the identified regions generally correspond to several large clustered ionized regions,
rather than a single ionized bubble. Altogether, the algorithm identifies $10$ ionized regions
across the MWA survey in the supercore configuration, but the identified regions have a lower overall quality than
in the $r^{-2}$ configuration. 

Our conclusion is that bubble detection is only marginally possible with the MWA-128. While the results
are unlikely to be very compelling, it is worth applying the matched filter to the first generation
surveys as an initial test. Even a few weakly identified candidate bubbles would provide compelling targets for follow-up
observations. Another possibility is to focus on {\em targeted searches} around known bright
sources for the MWA-128 (e.g., \citealt{Wyithe:2004ta}, \citealt{Friedrich:2012fy}).

\section{Favorable Antenna Configurations for Bubble Detection} \label{sec:LOFAR}

The possibility of imaging or identifying ionized regions from second
generation redshifted 21 cm surveys invites the question: how do
we optimize future surveys for this goal? It seems unlikely that
the optimal configuration for bubble detection is identical to that
for measuring the power spectrum, although power spectrum measurements
have mostly driven survey design considerations thus far.
In the case of the power spectrum, one
aims to minimize the error bar on power spectrum estimates in 
particular bins in wavenumber. The power spectrum error bar for
each $\k$-mode contains a thermal noise term and a sample variance (sometimes
called `cosmic variance') term. Because of the sample variance contribution,
the gain from reducing the thermal noise for a given $\k$-mode is limited: once
the thermal noise is reduced sufficiently far below
the sample variance, it is advantageous to instead measure a different
$\k$-mode on the sky within the $|\k|$ bin of interest. As a result, grouping
individual antennas into only small tiles to achieve a wide survey, generally
reduces the statistical error bars on power spectrum measurements compared
to antenna configurations with larger tiles that probe narrower fields of view.
For imaging and bubble detection, one aims for the best possible
signal to noise on {\em particular regions of the sky}. In other words, for good
imaging one wants to reduce the thermal noise to {\em well} below the sample variance level.
Grouping individual antennas into larger tiles, in order
to devote more collecting area to a narrower field of view, may be better
for this purpose.

In order to get some sense for these trade-offs, we consider here
a LOFAR-style interferometer with the specifications
listed in \cite{McQuinn:2006et}. Although the detailed
specifications for LOFAR have evolved somewhat (e.g., \citealt{Zaroubi:2012cy}),
(as have the MWA specifications), this is nonetheless a helpful case to consider. In particular,
our toy LOFAR-style interferometer has many fewer antenna tiles than the
MWA-500 but a substantially larger collecting area per tile. Specifically, the interferometer
considered has $N_{\text{a}} =
32$ antenna tiles, $A_{\text{e}} = 596\meter^2$ at our fiducial redshift (compared
to $A_{\text{e}} = 11.25 \meter^2$ for the MWA-500), $d_{\min}
= 100\meter$, and $d_{\max} = 2$ km. We assume that antenna tiles are
packed as closely as possible, consistent with $d_{\min} = 100 \meter$,
inside a compact core and that the
density subsequently falls off as $r^{-2}$, out to a
maximum radius of $r_{\max} = 1000 \meter$. These parameters are
meant to broadly represent an upgraded version of the existing LOFAR array, analogous to our
MWA-500 survey, which is an upgrade to the ongoing MWA-128 instrument. With 
these parameters, the LOFAR-style interferometer has more total
collecting area than the MWA-500 setup by a factor of a
few. 

\begin{figure}[h]
  \centering
  \includegraphics[width=8.4cm]{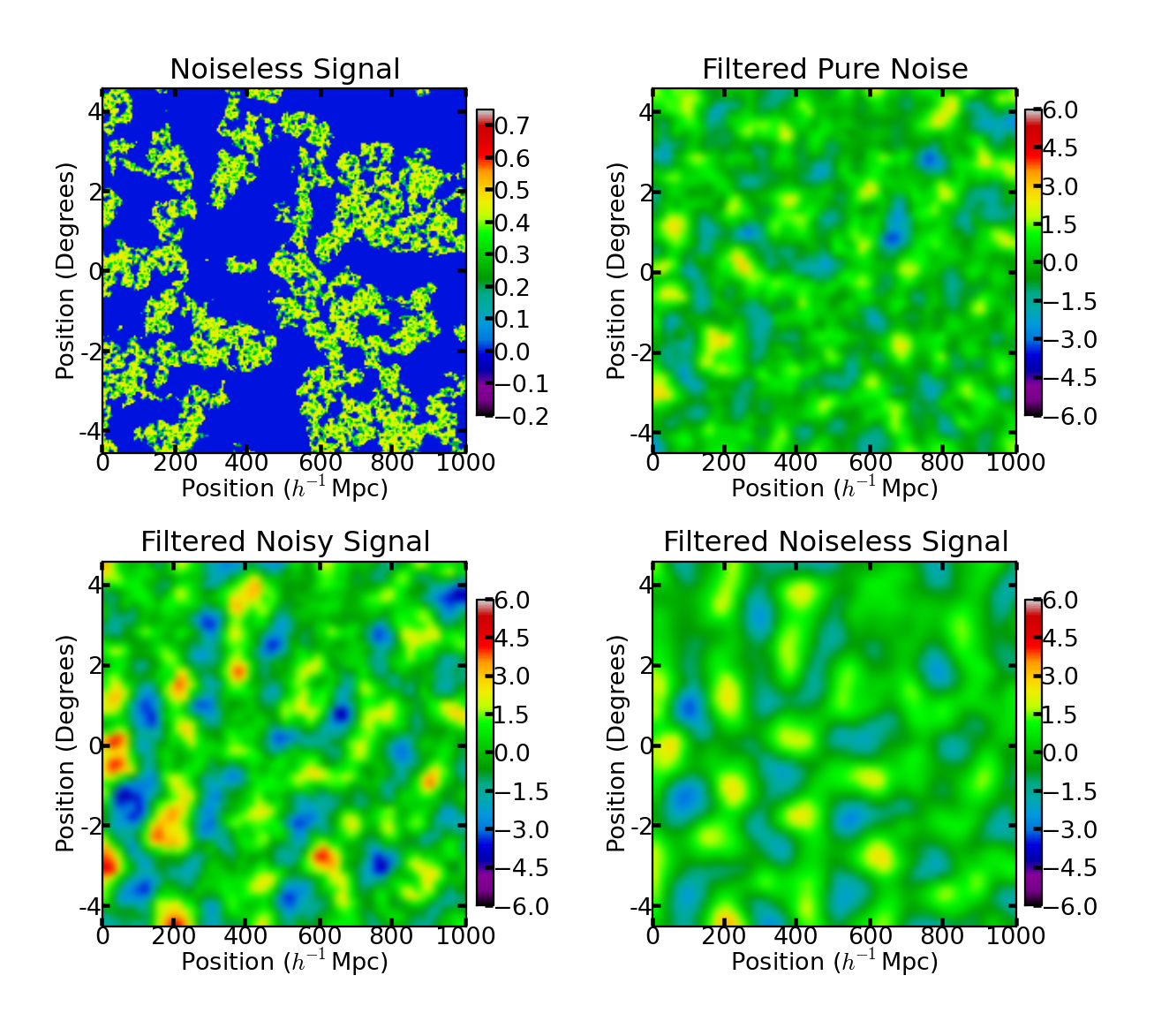}
  \caption{Bubble detection with a LOFAR-style interferometer. This figure is similar to Figure \ref{fig:ThreeFigureMatched}, except it is for the LOFAR configuration rather than the MWA-500. Additionally, all boxes in this figure have 
a side length of $426\hmpc$, corresponding to the field-of-view of the LOFAR-style
interferometer at $z = 6.9$.}
  \label{fig:ThreePanelLOFAR}
\end{figure}

The results of applying a matched filter to a data cube with simulated
LOFAR noise are shown in Figure \ref{fig:ThreePanelLOFAR}. Here we zoom in to show a portion of our simulation box that matches the smaller field of view of this LOFAR-like instrument. 
From the figure it
is evident that the filter removes large scale structures, a result
of the relatively large minimum baseline of this interferometer.
In addition, the maximum signal-to-noise achieved here is smaller
than with the MWA-500 (it drops from $10$ to $6.9$). Nonetheless, many
small-scale ionized regions in the unfiltered noise-less signal are
well preserved in the filtered noisy signal. This is consistent
with the idealized calculation of
Figure \ref{fig:ToySNR}, which showed that LOFAR should have
a higher signal-to-noise detection of small ionized regions, but
a reduced signal-to-noise otherwise. Because of this, the LOFAR-style
configuration is more successful during earlier stages of reionization
when the bubbles are still relatively small. In general, we find
that the LOFAR-style configuration detects slightly fewer bubbles overall
but with more success for $\avg{x_i} \lesssim 0.79$, while the MWA-500
has a greater level of success at later stages of the EoR.

This example suggests that the ideal configuration for bubble
detection is likely intermediate between the MWA-style and LOFAR-style
antenna configurations. It appears helpful to have more collecting
area on fewer baselines than the MWA, but a smaller minimum baseline
than in the LOFAR-style instrument is necessary to detect large
bubbles.  This deserves further study, however: for example, we have
neglected calibration requirements and systematic concerns. These
considerations will also certainly drive the experimental
design. As a further concrete example of how systematic concerns
could impact the design of future arrays, suppose 
foreground cleaning requires removing more
large scale modes than anticipated. In this case, it would make sense to focus
efforts on smaller bubbles.
This would shift the ideal configuration closer to a
LOFAR-style instrument with a larger minimum baseline.

\section{Comparisons to Previous Work} \label{sec:PreviousWork}

Previous work by \cite{Datta:2007nj} and \cite{Datta:2008ry} also
considered the possibility of detecting ionized regions in
noisy redshifted 21 cm data sets using a matched filter technique. 
The main difference between our study and this earlier work is
that these previous 
authors considered the prospects for detecting a {\em specific}
spherical ionized region of varying size, i.e., they considered the
detectability of
a spherical bubble at the origin, or offset slightly from the
origin. These authors also considered the case where the
bubble of interest was embedded in a variety of different ionization
environments; the bubble under consideration was not always isolated.
Altogether, their study is mostly similar to a targeted search, where one has a good prior
regarding the likely location of an ionized region. It also provides
a feasibility estimate for a more ambitious blind search. The main
advantage of a targeted search is that, if a region is known {\em a priori}
to be highly ionized, one need not worry about an entirely
false detection from a downward noise fluctuation. This then allows
a lower significance threshold for robust bubble detection, and may therefore be
the most feasible approach for the MWA-128 and other first generation
surveys. 

Nonetheless, our work is a significant extension to the earlier
work by \cite{Datta:2007nj} in that we conduct a blind search across
an entire mock survey volume. A detailed comparison with their
work is not straightforward given the difference between our approaches, but
both studies have a similar bottom-line conclusion: ionized regions
are detectable with surveys similar to the MWA-500.

\section{Conclusion} \label{sec:Conclusion}

We considered the prospects for making low-resolution images of the 21 cm sky
and for direct, blind detection of ionized regions using first and second generation
21 cm surveys. We find that a 500-tile version of the MWA, the MWA-500, is
potentially capable of detecting ionized regions. In our fiducial model, in
which $79\%$ of the volume of the Universe is ionized at $z_{\rm fid} = 6.9$,
the MWA-500 can find $\sim 150$ ionized regions in a $B = 6 \MHz$ chunk after $\sim 1,000$ hours of
observing time. First generation surveys, such as the MWA-128, are substantially
less sensitive. We find that the MWA-128 may, nonetheless, be able to detect a handful
of ionized regions across its survey volume, with $7$ expected in our fiducial model. The MWA-128 may be more effective at
identifying ionized bubbles using
targeted searches towards, for example, bright quasars (e.g. \citealt{Friedrich:2012fy}).

There are several possible future directions for this work. First, while we incorporate realistic levels
of thermal noise and mimic the effect of foreground cleaning, it will be important to
test the robustness of bubble detection with a more detailed model for foreground contamination,
and to consider systematic effects from calibration errors and the MWA instrumental response.
These considerations can also help in determining the optimal design for future surveys aimed at
bubble detection. Our first efforts considering which configurations of antenna tiles are favorable for bubble
detection, detailed in \S \ref{sec:LOFAR}, suggest that an observing strategy
intermediate to that of the MWA and LOFAR is favorable. It would also be interesting to
consider the prospects for bubble identification across a larger range of reionization models
than considered here. If the ionized regions at a given stage of reionization are, in fact, larger 
than in the models considered here, this should increase their detectability. On the
other hand, if the ionized regions are smaller than in our present models, this would likely
diminish detectability, at least for the MWA-500.

If blind bubble identification is indeed feasible in future 21 cm surveys, we believe
this will open up several interesting avenues of investigation. First, direct identification
of ionized regions can help to build confidence in early redshifted 21 cm detections.
Next, if the centers of ionized regions can be robustly identified, one may be able
to use the brightness temperature contrast between the signal near the bubble's center and its surroundings
to directly constrain the cosmic mean neutral fraction (e.g., \citealt{Petrovic:2010me}).
These authors also discuss how detected bubbles can be used to calibrate foreground
cleaning (\citealt{Petrovic:2010me}). Finally, identifying ionized regions in
redshifted 21 cm surveys allows one to commence follow-up observations, comparing galaxy properties inside detected bubbles
with those in more typical regions. Typical regions and likely neutral
regions can be identified as locations in the data cube with average and maximal signal-to-noise ratios, respectively, after
applying the matched filter.
Furthermore, if the edge of an ionized region can be identified
precisely enough, one might imagine targeted searches for galaxies at the edge of bubbles, close to neighboring neutral gas.
Spectroscopic
observations of these galaxies might then help to reveal 
the damping wing redward of the Ly-$\alpha$ line (e.g., \citealt{MiraldaEscude:1997qb}).
This would provide yet another means for
constraining the neutral fraction.

\section*{Acknowledgements} \label{sec:ThankYou}

We thank Judd Bowman and Piyanat Kittiwisit for related collaboration and feedback, and Matt
McQuinn for providing several helpful suggestions on a draft manuscript and for useful discussions. We also acknowledge a helpful report from the anonymous referee. 
We also thank Suvendra Dutta, Steve Furlanetto, Lars Hernquist, Peng Oh,
Jonathan Pritchard, Oliver Zahn, and Matias Zaldarriaga for discussions regarding imaging and bubble finding
in noisy 21 cm data.
MM and AL were supported by the NSF through grant
AST-1109156.


\bibliography{matt}

\end{document}